\documentclass[bibyear]{aa} 

\usepackage{graphicx}
\usepackage{txfonts}
\usepackage{hyperref}
\usepackage{natbib}
\usepackage{subcaption}
\usepackage{multirow}
\usepackage[normalem]{ulem}
\usepackage{booktabs}  
\usepackage{xcolor}

\begin{document}

   \title{Optimising the sample selection for photometric galaxy surveys}

   \author{M. Alemany-Gotor
          \inst{1,2}\thanks{E-mail: alemany@ice.csic.es}
          \and
          I. Tutusaus\inst{3}
          \and
          P. Fosalba\inst{1,2}
          }

   \institute{Institute of Space Sciences (ICE, CSIC), Campus UAB, Carrer de Can Magrans, s/n, 08193 Barcelona, Spain
         \and
             Institut d'Estudis Espacials de Catalunya (IEEC), Carrer Gran Capit\'a 2-4, 08034 Barcelona, Spain      
        \and
    Institut de Recherche en Astrophysique et Plan\'etologie (IRAP), Universit\'e de Toulouse, CNRS, UPS, CNES, 14 Av. Edouard Belin, 31400 Toulouse, France}

   \date{Received X, X; accepted X}


  \abstract
   {Determining cosmological parameters with high precision, as well as resolving current tensions in their values derived from low and high redshift probes, is one of the main objectives of the new generation of cosmological surveys. The combination of complementary probes in terms of parameter degeneracies and systematics is key to achieving these ambitious scientific goals.}
   {In this context, determining the optimal survey configuration for an analysis that combines galaxy clustering, weak lensing, and galaxy-galaxy lensing, the so-called 3x2pt analysis, remains an open problem. In this paper, we present an efficient and flexible end-to-end pipeline to optimise the sample selection for 3x2pt analyses in an automated way.}
   {Our pipeline is articulated in two main steps: we first consider a self-organising map to determine the photometric redshifts of a simulated galaxy sample. As a proof of method for stage-IV surveys, we use samples from the DESC Data Challenge 2 catalogue. This allows us to classify galaxies into tomographic bins based on their colour phenotype clustering. We then explore different redshift-bin edge configurations for weak lensing only as well as 3x2pt analyses in a novel way. Our method explores multiple configurations of perturbed redshift-bin edges with respect to the fiducial case in an iterative manner. In particular, we sample tomographic configurations for the source and lens galaxies separately.}
   {We show that, using this method we quickly converge into an optimised configuration for different numbers of redshift bins and cosmologies. Our analysis demonstrates that for stage-IV surveys an optimal tomographic sample selection can increase the figure of merit of the dark energy (DE) equation of state by a factor of $\sim$2, comparable to an effective increase in survey area of  $\sim$4 for non-optimal photometric survey analyses.}
  {}


   \maketitle
%

\section{Introduction}

In the current era of precision cosmology, surveys like {\it Euclid} (\cite{Euclid:2024yrr}) or the Legacy Survey of Space and Time~\citep[LSST;][]{2019ApJ...873..111I} from the Vera C. Rubin observatory are expected to provide data for more than a billion galaxies up to high redshift. In this scenario, the weak lensing (WL) and galaxy clustering (GC) probes, which measure excess correlations in the apparent shear and position of galaxies as a function of angular separation, respectively, will allow for unprecedented insight into the expansion history and growth rate of the large-scale structure (LSS) of the Universe. This will provide a significant improvement in cosmological parameter constraints as well as in the characterisation of DE. In addition, recent results from the Dark Energy Spectroscopic Instrument (DESI) collaboration \citep{DESI_Y2} have shown deviations from $\Lambda\text{CDM}$ of notable statistical significance in favour of dynamical DE. In light of this, there is increased interest in testing $w_{\rm 0} w_a\text{CDM}$ models with galaxy survey probes.\\

Since the LSS is sensitive to the influence of both dark matter and DE, the statistics of GC and WL allow for a mapping of the distribution of matter through space and time. GC requires some assumptions over the relation between dark and baryonic matter, and can only correctly trace dark matter as long as this relation, encompassed in the galaxy bias parameters, is fully characterised~\citep{galaxy_bias}. In contrast, WL is a direct tracer of dark matter. Combining the two probes, as well as galaxy-galaxy lensing (GGL), allows for a more robust analysis method by breaking degeneracies present in individual probes \citep{wl_degeneracies,Tutusaus2020}. Combining the three probes as well as their correlation is what is called the 3x2pt analysis, which
has been extensively used with current observations\,\citep{DESY3,KiDS1000} and will be used for stage-IV surveys \citep{Euclid_blanchard,LSST_3x2pt_optimization,2025AAS...24525015P}.\\

Optimising the sample selection for a single probe may not result in the optimal sample selection for a different probe. Moreover, in a realistic 3x2pt analysis the interplay between the different probes and their cross-correlations requires the optimisation to be done for the full set of probes involved. However, direct numerical attempts to find an optimised tomographic redshift bin configuration are hindered by the high computational cost of running a Bayesian inference pipeline at every step of the optimisation. \\

Several efforts have previously been performed in the literature to optimise the sample selection, some focussing on comparing equally-populated and equal-width tomographic redshift bins, e.g., \cite{andreas_paper,wong_3x2pt_optimization}, while others have presented more general optimisation methods limited to a lower number of redshift bins with lensing alone, e.g., \cite{rainbow}, \cite{Sipp_2020}. Several methods presented in \cite{LSST_3x2pt_optimization} attempt to improve the tomographic configuration by optimising the classification of galaxies into redshift bins. In this work we follow a different approach where we do not limit our exploration to a predefined set of possibilities (e.g., equally-populated vs. equal-width tomographic bins) and in this way we allow for different tomographic configurations to be tested. Nevertheless, we do not perform a completely general study of possible samples either, as that results in an intractable amount of possible configurations (e.g., for more than three tomographic bins a completely brute-force search of the optimal sample selection is already intractable, as shown in \cite{Sipp_2020}). Instead, we opt to optimise the tomographic bin edges separately for source and lens galaxy populations in an iterative manner, starting from the equally-spaced redshift bins case\footnote{We have checked that starting from equally populated z-bins we get consistent results}. For the proof of method presented in this paper we perform the optimisation for dynamic DE cosmology, using a tomographic 3x2pt analysis. Our implemented end-to-end pipeline is rather efficient, yielding a feasible computation time to converge into a robust solution ($\sim$1200 CPU hours). Our results show that using this method one can significantly improve the cosmological constraints (up to a factor of 2.5 in some cases) with respect to non-optimal tomographic analyses.\\

We have created a framework for tomographic optimisation that consists in an algorithm to run a pipeline of cosmological parameter inference based on the Cosmological Survey Inference System~{\citep[\texttt{CosmoSIS},][]{Zuntz_2015}}. We use as input data a photo-$z$ catalogue that we generated based on a simulation. The constraining power for each of the studied samples has been obtained using a Fisher matrix forecast technique.\\ 

The paper is organized as follows: in Sect.\,\ref{sec:photo-z} we describe the mock data and the method to obtain the photo-$z$ catalogue that we use as input. In Sect.\, \ref{sec:methodology} we present the formalism of the 3x2pt pipeline as well as the tomographic optimisation algorithm. In Sect.\,\ref{results} we show the results of the optimisation for two sets of cosmological parameters. In Sect.\,\ref{discussion} we discuss our results and compare them against those available in the literature. Finally, in Sect.\,\ref{conclusion} we present the main conclusions of the analysis.

\section{Mock input data}\label{sec:photo-z}

The method for tomographic optimisation introduced in this work requires an input photo-$z$ catalogue. As we will describe below, our input mock dataset does not include photo-z's, and thus we opted to generate a stage-IV-like photometric galaxy catalogue using a self-organizing map (SOM). For this purpose, we estimate photo-$z$'s based on each galaxy's ugrizy-band fluxes using a SOM. The SOM implementation closely follows a similar algorithm presented in \cite{rainbow}. Instead of using band fluxes directly, we work in colour space, defining the following colours: u-g, g-r, r-i, i-z, z-y.\\

\subsection{Simulated data}

The data that we have used comes from the DESC Data Challenge 2~\citep[{DESC DC2},][]{DESCDC2}, and consists in a simulated galaxy catalogue mimicking a small patch of the expected observations of the future Rubin Observatory. We downloaded our samples using CosmoHub \citep{TALLADA2020100391}, \citep{2017ehep.confE.488C}. DESC DC2 is based on the CosmoDC2 simulation~\citep{LSSTDarkEnergyScience:2019hkz}, \citep{DESCDC2}. It covers $\sim 440$ squared degrees with photometric data of galaxies up to redshift 3 in the ugrizy bands. The simulation was created as part of the preparation for {LSST} \citep{LSST_3x2pt_optimization,3x2pt_optimization_2}, which is expected to provide an unprecedented amount of astronomical data that will enable scientists to address a wide range of critical questions in observational cosmology, constraining the nature of DE, dark matter, and providing a better understanding of how galaxies form.\\

The {DESC DC2} simulation includes several key features that make it an ideal tool to study these phenomena. First, it includes a large number of galaxies, modelled using sophisticated algorithms that take into account a wide range of physical processes, such as star formation, gas dynamics, and feedback from supernovae and black holes. Second, the simulation covers a large volume of space, spanning over 440 square degrees on the sky, with a high level of detail and resolution. The CosmoDC2 catalogue is based on the Outer Rim simulation, an N-body halo simulation by \cite{cosmodc2_halosim} which assumes a $\Lambda\text{CDM}$ model with the cosmological parameters being the matter density $\Omega_{{\rm m}} = 0.27$, the baryonic matter density $\Omega_{{\rm b}} = 0.045$, the spectral index $n_{{\rm s}} = 0.963$, the dimensionless Hubble parameter $h = 0.71$, the amplitude of matter fluctuations in spheres of 8\,Mpc$/h$ $\sigma_8 = 0.8$. Since it assumes $\Lambda \text{CDM}$, the values for the DE equation of state parameters are $w_{\rm 0} =-1$ and $w_a=0$. It consists of a 4.225 Gpc$^3$ box with a $10.240^3$ particles and a resolution of ~$2.6\times 10^9M_{\odot}$, producing 101 snapshots from a starting redshift of $z=10$ to $z=0$.\\

We assume the simulation cosmology when performing our analysis, 
except for the addition of dynamical DE with the {$w_{\rm 0} w_a\text{CDM}$} model, which includes the following dynamical DE modelling with a varying equation of state parameter:

\begin{equation}
    w(z) = w_{\rm 0} + w_a \frac{z}{1+z}\,{,}
\end{equation}
where $w_{\rm 0}$ corresponds to the value of the equation of state parameter today and $w_a$ parametrises the variation in the dynamics of DE over redshift \citep{dynamical_dark_energy,dynamical_dark_energy_2}. In this context, the optimisation we perform is to minimise the errors around the reference cosmology of the simulated data we use, $\Lambda\text{CDM}$, across the {$w_{\rm 0} w_a\text{CDM}$} model parameters.\\

The catalogue contains the fluxes, positions and shear of the galaxies. We use the colours defined as the differences in magnitude of pairs of fluxes, propagating the flux errors into colour errors. The photo-$z$ estimation for each galaxy is performed using the galaxies' colour vector and associated error.

\subsection{Photo-z estimation}

A SOM is essentially a method to reduce the dimensionality of high-dimensional data by projecting it onto a lower-dimensional representation. It consists of a 2-dimensional grid map composed of $n_1 \times n_2$ discrete points called neurons or voxels. Each voxel $(j,k)$ in this grid has an associated $m$-dimensional weight vector $\vec{w}_{jk}$, where the indices $jk$ specify the voxel's position on the map. In particular, the weights will have 5 dimensions corresponding to the 5 colour values that each galaxy has. Initially, these weight vectors are set to random values. During the training process, the SOM algorithm iteratively adjusts these weights so that voxels with similar weight vectors become clustered together on the map. This creates a topology where galaxies with similar colour phenotypes will be mapped to nearby regions.\\

Once training is complete, individual galaxies are assigned to specific voxels based on the best match between the galaxy's observed colours and the weight vectors of the voxels, typically using a minimum distance criterion. The photo-$z$ estimation is then performed by assigning to each galaxy a redshift value using the $n(z)$ distribution associated with its matched voxel. The $n(z)$ of a given voxel is composed of the spectroscopic redshift of the training galaxies that were assigned to that voxel. Thus, the SOM reduces the dimensionality of our data and allows for both photo-$z$ estimation and for the possibility of selecting phenotypically similar galaxies based only on their proximity on the map. More details on the specific implementation of the SOM are given in Sect. \ref{appendix:som} in the appendix.\\

\subsubsection{Training and photo-$z$ galaxy catalogues}

For our training sample we have selected an r-band magnitude-limited ($<24.0$) sample in a patch of 4 squared degrees, containing $\sim 10^5$ galaxies. The patch size has been selected so that the number of galaxies was similar to the sample used for the training of a SOM in \cite{Masters:2015asa}. Our photo-$z$ galaxies are selected by an r-band magnitude cut at $<24.5$ leading to a galaxy sample that contains $\sim 3\times 10^7$ photometric galaxies over a patch of 225 squared degrees. The magnitude cut for the photometric galaxy sample was chosen to resemble stage-IV forecasts \citep{andreas_paper}. The magnitude cuts of the training sample are optimistic and motivated by the need to achieve a meaningful level of high-redshift galaxies. This is because the ultimate goal of our work is to present a tomographic sample optimisation method for a stage-IV-like photometric galaxy sample and our training galaxy sample was adjusted accordingly. Finally, we also selected galaxies with true redshift below $z=2$ for both catalogues. This is because introducing galaxies above that redshift will worsen the general performance of the SOM without providing an adequate characterization of those high redshift galaxies. We note however that this selection should not limit the general validity of our approach for stage-IV experiments, as we cover a redshift range comparable to current or future surveys. We present a quantitative analysis of the performance of the photo-$z$ estimation using the SOM in Sect.\,\ref{photo-z estimation:performance}.\\

\subsubsection{Performance evaluation}\label{photo-z estimation:performance}

\begin{figure}
  \centering

  \begin{subfigure}{0.45\textwidth}
    \centering
    \includegraphics[width=\columnwidth]{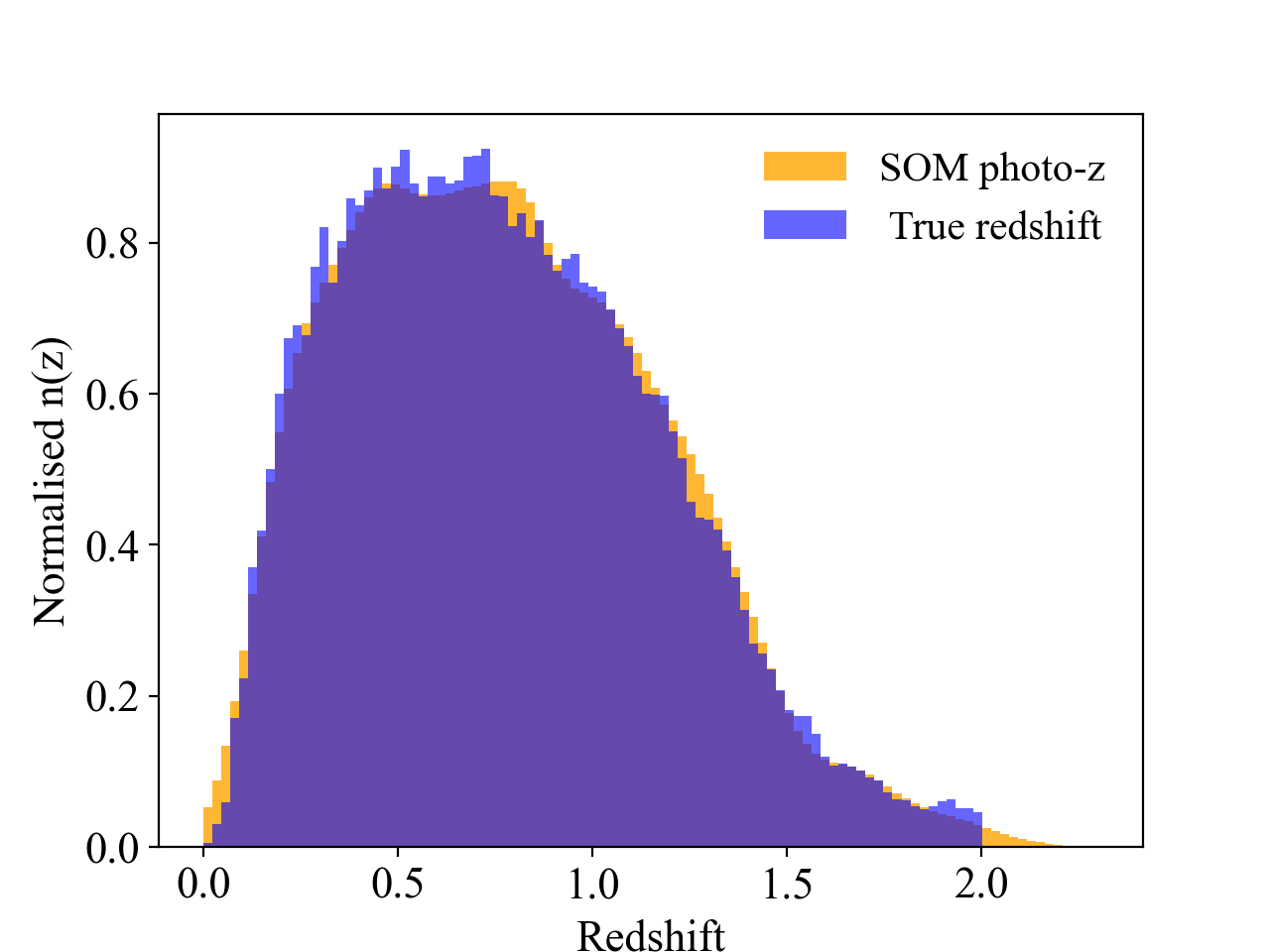}
 \caption{Normalised distributions of the true redshift of galaxies from the simulated catalogue (purple) and of the estimated photo-$z$ (orange).}
 \label{fig:som_nofz}
  \end{subfigure}
  \hfill
  \begin{subfigure}{0.45\textwidth}
    \centering
    \includegraphics[width=\columnwidth]{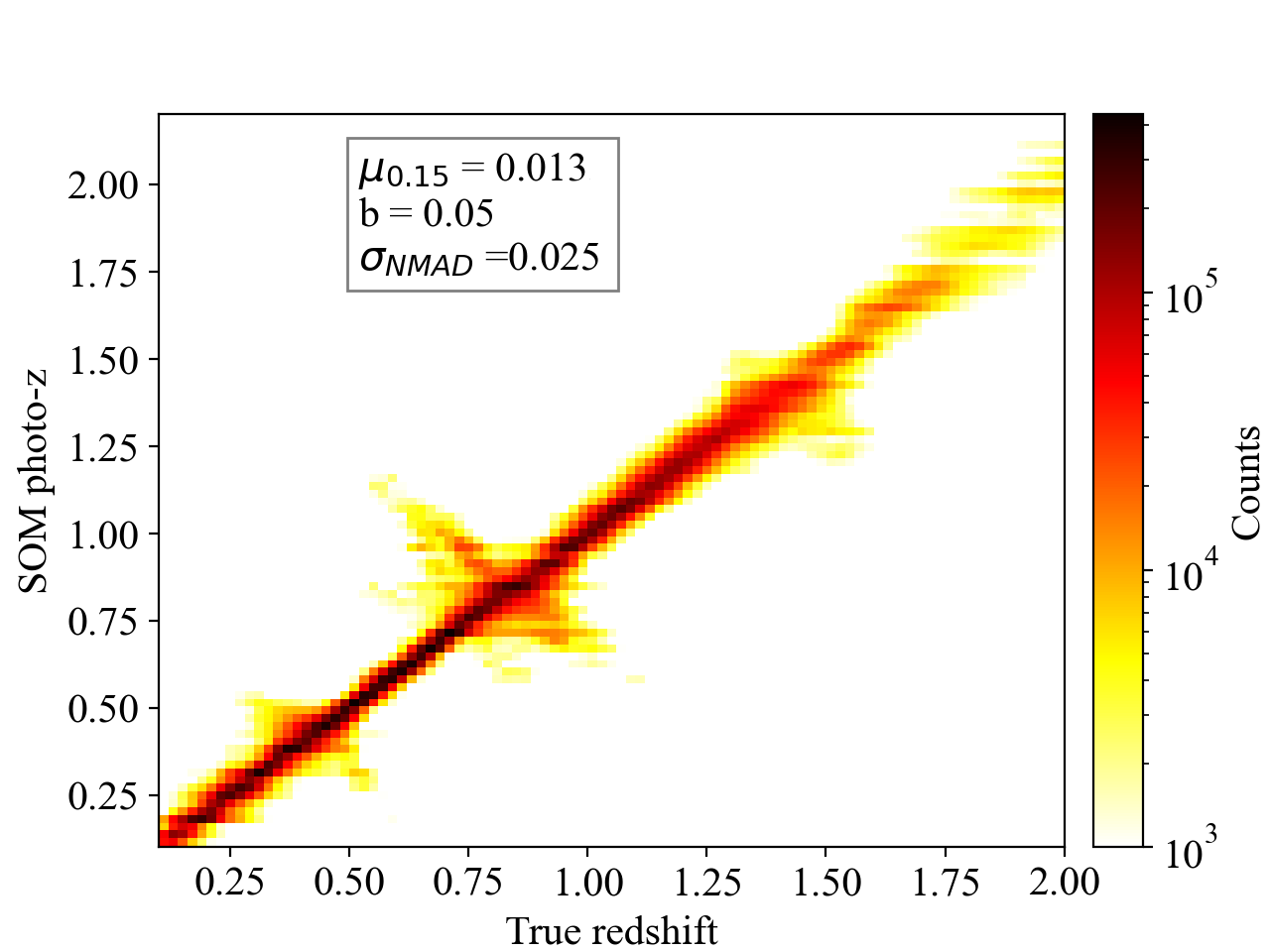}
 \caption{Performance of the SOM in the estimation of the photo-$z$'s of galaxies.}
 \label{fig:true-photo-z}
  \end{subfigure}
    \caption{SOM performance evaluation.}
 \label{fig:SOM_performance}
\end{figure}

We evaluate the performance of the SOM in properly assigning photo-$z$'s to galaxies based on their colours. Since the data that we use have been generated with a simulation, we know the true redshift of the galaxies in the catalogue. This allows for a precise characterisation of the performance of the photo-$z$ estimation method. We also compare the distributions of the true redshift and of the SOM-estimated photo-$z$ in Fig.\,\ref{fig:som_nofz}, where the photo-$z$ distribution closely matches the original. The number of galaxies at high redshift in the training sample is low, leading to poorer photo-$z$'s for high-redshift galaxies. The SOM photo-$z$ $n(z)$ resembles the true redshift $n(z)$ nonetheless, even at high redshift.\\

In Fig.\,\ref{fig:true-photo-z} we evaluate the performance of the SOM through the use of three metrics: the outlier rate $\mu_{0.15}$, the bias $b_z$, and the normalised median absolute deviation $\sigma_{\text{NMAD}}$. These metrics together provide an assessment of the accuracy of the photo-$z$ estimates produced by the SOM. The $\mu_{0.15}$ metric quantifies the fraction of so-called catastrophic outliers and is a measure of the representativeness of the spectroscopic sample. Catastrophic outliers occur mostly due to degeneracies in the relation between colour and redshift of galaxies, where the spectral energy distribution is compatible with multiple voxel colour profiles. It is defined as:

\begin{equation}
    \mu_{0.15} = \frac{1}{N} \sum_{i=1}^N \left(|z_{\text{ph}}-z_{\text{sp}}| > 0.15(1+z_{\text{sp}}) \right)\,,
\end{equation}
where $N$ is the number of galaxies, $z_{\text{ph}}$ is the photometric redshift and $z_{\text{sp}}$ is the spectroscopic redshift (in our case, the true redshift of the galaxy). Galaxies whose assigned voxel is empty are discarded, removing misfit galaxies that could result in catastrophic outliers. This results in 5\% of the galaxies in the photometric catalogue being discarded. The percentage of galaxies that are discarded depends on the representativeness of the training sample. A less representative training sample $n(z)$ will result in a higher percentage of discarded galaxies.\\

The normalised median absolute deviation ($\sigma_{\text{NMAD}}$) gives us an estimate of the dispersion of the sample that is less sensitive to catastrophic outliers. It primarily quantifies the quality of the photometric data rather than the representativeness of the sample. It is defined like:

\begin{equation}
    \sigma_{\text{NMAD}} =1.4826\,\times \, \text{median} \left(|z_{\text{ph}}-z_{\text{sp}}|- \text{median}(|z_{\text{ph}}-z_{\text{sp}}|) \right)\,.
\end{equation}

The $b_z$ bias of the photo-$z$ catalogue measures possible systematic offsets in the estimation. It does so by measuring the slope between the true redshift and the SOM photo-$z$'s. Different possibilities can result in a non-zero $b_z$. Since our simulated galaxy catalogue is limited to $z<2$ but our photo-$z$'s can be estimated to be slightly higher because of the use of Gaussian fits to the $n(z)$ of a given voxel, we can expect a slightly positive bias. The bias is defined as:

\begin{equation}
    b = \left\langle z_{\text{ph}} -z_{\text{sp}}\right\rangle\,.
\end{equation}

An outlier rate of $\mu_{0.15}= 0.013$ and $\sigma_{\text{NMAD}} = 0.025$ is similar to other photo-$z$ catalogues used in tomographic configuration analysis, e.g., the optimistic case in \cite{andreas_paper}. We also expected a slightly positive bias $b_z$, which we do find. Our estimated photo-$z$ galaxy catalogue is thus similar to other stage-IV forecasting work.\\

\section{Methodology}\label{sec:methodology}

\subsection{\texttt{CosmoSIS} pipeline of cosmological parameter inference}

Our pipeline of cosmological parameter inference is based on the \texttt{CosmoSIS} framework. In this section we detail the configuration of the pipeline and outline the modelling of the observables as well as the systematic effects. We closely follow the public DES-Y3 pipeline \citep[more details in][]{DESY3}, using a similar treatment of non-linearities and imposing comparable priors on our nuisance parameters.\\

Our matter power spectrum has been generated using {\texttt{CAMB}} (\cite{CAMB}, \cite{CAMB_1}) as implemented in the corresponding {\texttt{CosmoSIS}} module, with the non-linear power spectrum being generated using the Halofit~\citep{halofit_takahashi} recipe as implemented in {\texttt{CAMB}}.\\

The matter power spectrum has then been used to generate the predictions of the three data vectors of 3x2pt: the angular power spectrum for cosmic shear ($C_{\ell}^{\gamma\gamma}$), galaxy clustering ($C_{\ell}^{gg}$){,} and {their} cross-correlation ($C_{\ell}^{g\gamma}$). We compute the values for each $C_\ell$ in 60 logarithmically-spaced bins from $\ell =10 $ to $\ell =1500$, which is in line with the pessimistic case presented in~\cite{Euclid_blanchard}. We go beyond the Limber approximation between multipoles 10 and 200 by following the approach from \cite{Fang:2019xat} as implemented in \texttt{CosmoSIS}, which also includes redshift space distortions (RSD). RSD are a source of systematic uncertainty resulting from the effect of peculiar velocities of galaxies on their measured or estimated redshift and we include them in the prediction of the galaxy clustering data vectors, where the effect is most significant. The power spectra under the Limber approximation are given by:

\begin{equation}
C_{\ell}^{\gamma\gamma} = \int_0^{\chi_H} d\chi\,\frac{q_{\gamma\gamma}^2(\chi)}{\chi^2}P_{\rm m}\left(\frac{\ell+1/2}{\chi},\chi\right)\,,
\end{equation}

\begin{equation}
C_{\ell}^{gg} = \int_0^{\chi_H} d\chi\,\frac{q_{gg}^2(\chi)b^2(\chi)}{\chi^2}P_{\rm m}\left(\frac{\ell+1/2}{\chi},\chi\right)\,,
\end{equation}

\begin{equation}
C_{\ell}^{\gamma g} = \int_0^{\chi_H} d\chi\,\frac{q_{\gamma g}(\chi)b(\chi)}{\chi}P_{\rm m}\left(\frac{\ell+1/2}{\chi},\chi\right)\,,
\end{equation}
where $P_{\rm m}$ is the matter power spectrum, $b(z)$ is the linear galaxy bias, $\chi$ is the comoving angular diameter distance, and $\chi_H$ is the comoving distance to the horizon. The kernel functions, $q(\chi)$ are defined as follows:

\begin{equation}
    q_{\gamma\gamma}(\chi) = \frac{3 H_0^2\Omega_{{\rm m}}}{2c^2 a(\chi)} \int_\chi^{\chi_h} d\chi' n_{\gamma} (\chi') \chi \left(1-\frac{\chi}{\chi'} \right)\,,
\end{equation}

\begin{equation}
    q_{gg}(\chi) = n_g(z) \frac{dz}{d\chi}\,,
\end{equation}

\begin{equation}
    q_{\gamma g}(\chi) = \frac{3 H_0^2\Omega_{{\rm m}}}{2c^2 a(\chi)} \int_\chi^{\chi_h} d\chi' n_{\gamma} (\chi') \left(1-\frac{\chi}{\chi'} \right)\,,
\end{equation}

where $n_\gamma (\chi)$ and $n_g (\chi)$ are the redshift distributions of the source and lens galaxies respectively. 

\subsubsection{Systematic effects}\label{section:systematics}

In order to achieve our goal of optimising the tomography for a realistic stage-IV-like survey we have to implement the appropriate modelling of systematic effects into our pipeline. In this section we outline our treatment of systematic effects. Since the nuisance parameters of the source and lens galaxies distributions as well as the galaxy bias and shear calibration parameters depend on the galaxy distributions, we compute them for each tomographic realisation (both for the lens and the source samples).\\

We employ a linear galaxy bias model in our analytic data vectors. For a sufficiently realistic estimation of the fiducial values of the galaxy bias, we used the polynomial fit presented in \citet{Euclid:2021rez}, which is representative of a stage-IV survey like \textit{Euclid}. The estimation of the fiducial galaxy bias in each bin is given by:

\begin{equation}
    \text{b}(z_{\text{avg}}) = 0.5125 + 1.377\,z_{\text{avg}} + 0.222\,z_{\text{avg}}^2  -0.249\,z_{\text{avg}}^3\,,
\end{equation}
with $z_{\text{avg}}$ being the average redshift of the galaxies in the tomographic bin.
\\

On the other hand, weak lensing measurements are affected by the so-called intrinsic alignment (IA) of galaxies, whereby gravitational effects result in a correlation in the orientation of galaxies due to tidal gravitational fields. This correlation can mimic or contaminate the shear signal produced by gravitational lensing. It is thus necessary to take IA into account in our modelling to avoid a bias in our cosmological constraints \citep{Heavens:2000ad}.\\

We use the tidal-alignment tidal-torque model (TATT) \citep{Blazek:2017wbz} as implemented in the \texttt{CosmoSIS} framework in order to model for IA effects. The TATT model is a perturbative expansion of the linear density field to the second order, with the intrinsic shear of galaxies being described by:

\begin{equation}
    \gamma^{\text{IA}}_{ij} = C_1 \, s_{ij} + C_2 \left(s_{ik} s_{kj} -\frac{1}{3} s^2\right) + C_{1\delta} \left(\delta s_{ij} \right)\,,
\end{equation}
where $s_{ij}$ is the tidal tensor and $C_1$ and $C_2$ are defined as:

\begin{equation}
    C_1(z) = A_1\bar{C_1}\rho_{crit} \frac{\Omega_{{\rm m}}}{D(z)}\left(\frac{1+z}{1+z_0}\right)^{\alpha_1}\,,
\end{equation}

\begin{equation}
    C_2(z) = 5\,A_2\bar{C_1}\rho_{crit} \frac{\Omega_{{\rm m}}}{D^2(z)}\left(\frac{1+z}{1+z_0}\right)^{\alpha_2}\,,
\end{equation}
with $\rho_{crit}$ being the critical density of the Universe, $D(z)$ being the linear growth function, $z_0$ being the pivot redshift and the free parameters of the model being $A_1$, $A_2$, $\alpha_1$, and $\alpha_2$. The term $C_{1\delta}$ is related to $C_1$ through the tidal alignment bias, $b_{\rm TA}$, like $C_{1\delta} = b_{\rm TA}\, C_1$. This model captures the complexity of IA for galaxies with different morphologies while also aiding in the modelling of IA dynamics at smaller and more non-linear scales.\\

\begin{table}

\caption{Fiducial values for the systematic effects in our IA modelling, shear calibration, and $n(z)$ errors. The step sizes for the numerical derivatives are discussed in Sect. \ref{sec:fisher}.}
\begin{tabular}{lll}
\multicolumn{1}{l|}{Parameter} & \multicolumn{1}{l|}{Fiducial value} & Step size \\ \hline
\multicolumn{3}{l}{Cosmological parameters}                                           \\ \hline
$w_{\rm 0}$                          & -1.0                                 & $1.6\times10^{-2}$           \\
$w_a$                          & 0.0                               & $1.6\times10^{-2}$           \\
$\Omega_{{\rm m}}$                     & 0.27                                & $4.2\times10^{-3}$           \\
$\Omega_{\rm b}$                     & 0.045                                & $7.2\times10^{-4}$           \\
$n_{\rm s}$                        & 0.96                                 & $1.5\times10^{-2}$            \\
$h$                          & 0.71                                & $1.1\times10^{-2}$                     \\ 
$\text{ln} (10^{10}\text{A}_{\rm s})$                        & 3.0                                 & $4.8\times10^{-2}$            \\ \hline
\multicolumn{3}{l}{Intrinsic alignment parameters}                                           \\ \hline
$A_1$                          & 0.7                                 & $2.0\times10^{-1}$           \\
$A_2$                          & -1.4                               & $2.0\times10^{-1}$           \\
$\alpha_1$                     & -1.7                                & $2.0\times10^{-1}$           \\
$\alpha_2$                     & -2.5                                & $2.0\times10^{-1}$           \\
$b_{\rm TA}$                        & 1.0                                 & $4.0\times10^{-2}$            \\
$z_{0}$                          & 0.62                                & -                     \\ \hline
\multicolumn{3}{l}{Shear calibration parameters}                                             \\ \hline
m\_i                           & 0.0                                 & $4.0\times10^{-3}$           \\ \hline
\multicolumn{3}{l}{Lens photo-z errors}                                                      \\ \hline
$\Delta z_i$                   & 0.0                                 & $2.0\times10^{-3}$         \\
$\sigma_i$                     & 1.0                                 & $1.6\times10^{-2}$            \\ \hline
\multicolumn{3}{l}{Source photo-z errors}                                                    \\ \hline
$\Delta z_i$                   & 0.0                                 & $4.0\times10^{-3}$          \\ \hline
\multicolumn{3}{l}{Galaxy bias}                                                    \\ \hline
$b_i$                   & By Eq. (7)                                 & $1.6\times10^{-2}$\,$b_i$          \\ \hline
\multicolumn{3}{l}{Magnification bias}                                                    \\ \hline
$s_i$                   & By Eq. (14)                                 & $1.6\times10^{-2}$\,$s_i$          \\ 
\end{tabular}\\

\label{table:TATT_table}
\end{table}

We have also considered a shear multiplicative bias, $m_i$, to account for uncertainties in the shear calibration that takes the form:

\begin{equation}
    \hat{\gamma}_i = (1+ m_i) \gamma_i\,,
\end{equation}
where $i$ denotes the bin index, $\gamma_i$ being the true angular correlation function and $\hat{\gamma}_i$ being the observed angular correlation function. This correction is necessary because small errors in the shape measurement can systematically overestimate or underestimate the amplitude of the shear signal.\\

The rest of the systematic effects are related to the photo-$z$ distribution for the lens and source galaxy redshift bins and account for either deviations in the mean of the distribution of both lens and source bins or for deviations in the width of the distributions of lens galaxies. This procedure is similar to \cite{DES:2021bpo}, where only a shift in the mean of the source galaxies is considered. This is because the uncertainties in higher order modes of the source $n(z)$'s besides the mean are negligible.\\

Another source of systematic error that we model in our pipeline is magnification. Magnification is a gravitational lensing effect by which the apparent size and the flux of source galaxies are modified. This distortion can affect clustering measurements and bias the angular power spectra of GC and GGL. We include modelling of magnification as implemented in \texttt{CosmoSIS}, where the GC and GGL angular power spectra are, respectively:

\begin{equation}
    C_{\ell}^{gg,\text{obs}} = C_{\ell}^{gg}  +\left( 5s_i -2 \right)^2 C_{\ell}^{\kappa\kappa} + 2\left( 5s_i -2 \right) C_{\ell}^{g\kappa}\,,
\end{equation}

\begin{equation}
    C_{\ell}^{g\kappa',\text{obs}} = C_{\ell}^{g\kappa'}  +\left( 5s_i -2 \right)C_{\ell}^{\kappa\kappa'}\,,
\end{equation}
where $\kappa$ refers to the magnification field, $C_{\ell}^{\kappa\kappa}$ is the angular power spectra for the magnification-magnification correlation and $C_{\ell}^{g\kappa}$ is the galaxy count-magnification correlation. Similarly to the case for the galaxy bias, the estimation of the fiducial magnification bias of each redshift bin is based on a fit based on Euclid mission-like simulations from \cite{Euclid:2021rez}:

\begin{equation}
    s_i(z) = 0.0842 + 0.0532\, z_{\text{avg}} + 0.298\,z^2_{avg}  -0.0113\,z^3_{avg}\,,
\end{equation}  
with $z_{\text{avg}}$ being the average photometric redshift of the galaxies in a given tomographic redshift bin.\\

\subsubsection{Covariance matrix estimation}

We consider a Gaussian covariance for our observables and generate it with the implementation in \texttt{CosmoSIS}. Since the purpose of our work is the relative improvement in the constraining power of a given sample selection choice, we do not consider higher-order terms like non-Gaussianities or super-sample covariance effects in our forecast, for simplicity. The equations to model our covariance are as follows:

\begin{align}
     &\textbf{Cov} \left(C(\ell)^{AB}_{ij}, C(\ell')_{kl}^{A'B'}\right) = \frac{1}{(2\ell+1)\,f_{\text{sky}} \Delta \ell} \cdot \notag\\ 
    &\left[(C(\ell)_{ik}^{AA'} + N_{ik}^{AA'})(C(\ell')_{jl}^{BB'} + N_{jl}^{BB'}) + \right. \notag\\
    & \left.+ (C(\ell)_{il}^{AB'} + N_{il}^{AB'})(C(\ell')_{jk}^{BA'} + N_{jk}^{BA'})\right] \delta_{\ell \ell'}\,,
\end{align}
where $A$ and $B$ go through the probes of GC, WL, and GGL and the indices $i$, $j$, $k$, and $l$ refer to the tomographic bins. The $N^{AB}_{ij}$ describes the noise terms, which are the following:

\begin{equation}
    N^{GC}_{ij}= \frac{\delta^K_{ij}}{\bar{n_i}}\,,
\end{equation}

\begin{equation}
    N^{WL}_{ij}= \frac{\sigma^2_\epsilon \delta^K_{ij}}{\bar{n_i}}\,,
\end{equation}

\begin{equation}
    N^{GGL}_{ij}= 0\,,
\end{equation}
where $\bar{n}_i$ is the density of galaxies per unit of area of the sky for a given tomographic bin and $\sigma_\epsilon = 0.3$ is the ellipticity dispersion of the galaxies in our catalogue, chosen to be similar to comparable forecasts (e.g., \cite{Euclid_blanchard}). The density of galaxies per bin will depend on the way in which the tomographic configuration is defined, with the total density of galaxies being $\approx 11\,\rm gal/arcmin^2$. We set the fraction of observed sky at $f_{sky}=0.367$, illustrative of stage-IV experiments (e.g, Euclid). The number density depends mostly on the magnitude cut for the training sample of the SOM. Fainter magnitude cuts will result in a larger number density but introduce an unrealistic characterisation of high-redshift galaxies. \\

\subsubsection{Fisher formalism}\label{sec:fisher}
In order to forecast the constraints on cosmological parameters under various configurations of analysis we employ the Fisher formalism, as has been done in other works with forecasting for stage-IV surveys (e.g., \cite{LSST_3x2pt_optimization}, \cite{Euclid_blanchard}). The Fisher matrix quantifies the information on cosmological parameters given a set of observables after imposing several assumptions. One of which consists in assuming Gaussian posteriors and then sampling the gradient along an axis. This allows for the estimation of the expected uncertainties in the parameters as we test different tomographic configurations. In the case of LSS observations, the Fisher matrix can be used to estimate the expected constraints on the parameters of the $w_{\rm 0} w_a\text{CDM}$ model given a set of data vectors for 3x2pt.\\

Given an observed dataset $Q$ with a likelihood function $L(Q|\theta)$ depending on a set of parameters $\theta$, the Fisher matrix $F_{\alpha \beta}$ for a pair of parameters $\theta_\alpha$ and $\theta_\beta$ is defined as:

\begin{equation}
F_{\alpha \beta} = - \left\langle \frac{\partial^2 \ln L}{\partial \theta_\alpha \partial \theta_\beta} \right\rangle\,,
\end{equation}
where $\ln L$ is the natural logarithm of the likelihood function. The inverse of the Fisher matrix, $F^{-1}_{\alpha \beta}$, is an estimate of the covariance matrix of the parameters.  

For our case with 3x2pt, which includes WL, GC, and GGL, the Fisher matrix is given in terms of the previously-defined covariance as: 

\begin{align}
     &F_{\alpha \beta} = \sum_{\ell_{min}}^{\ell_{max}} \sum_{ABA'B'}\sum_{ijkl} \notag \\ 
    &\,\,\,\,\,\,\,\,\,\,\,\,\,\,\,\frac{\partial C_{ij}^{AB}(\ell)}{\partial \theta_\alpha} \textbf{Cov} \left(C(\ell)^{AB}_{ij}, C(\ell)_{kl}^{A'B'}\right) \frac{\partial C_{kl}^{A'B'}(\ell)}{\partial \theta_\beta}\,,
\end{align}
where the indices $ABA'B'$ represent the probes, $ijkl$ represent the tomographic bins, and $\theta_\alpha \theta_\beta$ represent the parameters of the model.\\

We use \texttt{CosmoSIS} to compute the Fisher matrix numerically with the covariance matrix computed with the same code. The resulting Fisher matrix is used to estimate the expected constraints on the cosmological parameters of interest and to determine the optimal tomographic configuration for different scientific goals. We do that by quantifying the information about a given set of parameters using the commonly-used metric called figure of merit (FoM), as defined in \cite{Wang:2008zh}, \cite{fom_ref}

\begin{equation}
    \text{FoM} = \sqrt{\det{F_{\alpha \beta}}}\,,
\end{equation}
where $F_{\alpha \beta}$ is the marginalised Fisher sub-matrix for a set of cosmological parameters. In this way, we can quantify, using a single value, the improvement in the constraints after the sample selection optimisation is performed. In order to compute the Fisher matrix, we evaluate numerical derivatives of the observables with respect to each parameter. These derivatives are calculated using finite differences and the step sizes for each parameter are described in Table \ref{table:TATT_table}.\\

\subsection{Tomographic optimisation}\label{sample optimization}

In order to construct the tomographic bins that will be used as the source and lens samples for 3x2pt, we select groups of voxels according to a given criterion and define the bins with all the galaxies that had been assigned to each of the voxels during the photo-$z$ estimation. Our method uses photo-$z$'s as the quantity by which to classify voxels into tomographic bins. Trying to find the optimal configuration by selecting any possible voxel combination is computationally intractable, even when using hyperparameters to reduce the dimensionality of the selection \citep[e.g.,][]{Sipp_2020}, thus alternative methods to quickly test and optimise tomographic configurations are needed.\\

\begin{figure}
  \centering
    \includegraphics[width=\linewidth]{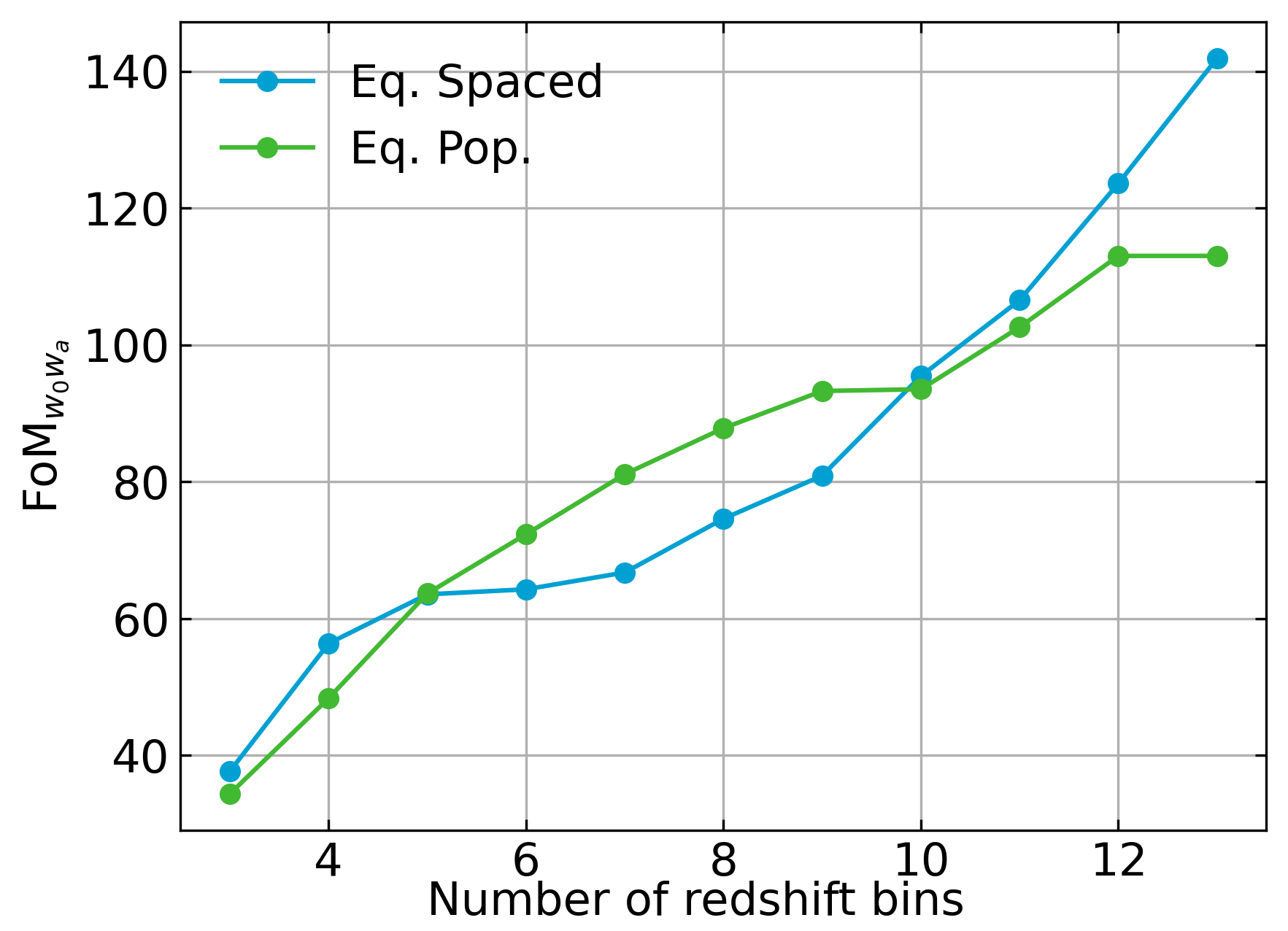}
    \caption{FoM 
    for $w_{\rm 0} w_a$ as we increase the number of redshift bins generated by assigning galaxies to the equally-spaced or equally-populated redshift bins.}
 \label{fig:fiducial}
\end{figure}

We select the redshift bins by defining non-overlapping redshift ranges and grouping all the voxels with an average redshift within that range. The tomographic configuration to be used for stage-IV surveys is still an unsettled issue and surveys like KiDS-1000 \citep{KiDS:2020suj} or the Hyper Suprime-Cam \citep{Dalal:2023olq} have used equally-spaced redshift bins, while DES-Y3 \citep{DESY3} or KiDS Legacy \citep{Wright:2025xka} used equally-populated bins. The best-performing tomography between equally-populated and equally-spaced redshift bins will depend on factors like the number of bins and survey characteristics like the redshift range. In Fig.\,\ref{fig:fiducial} we present the performance of these two configurations measured with the FoM for the set of the two DE equation of state parameters. Equally-populated redshift bins outperform equally-spaced bins for an intermediate number of redshift bins, while equally spaced bins are best for a higher number of bins, similarly to the results by \cite{andreas_paper}.\\

Our method to improve the selection of redshift bins takes the equally-spaced case as a starting point for the tomography optimisation\footnote{But we note that we get consistent results when starting from equally-populated bins, as shown in Sect. \ref{sec:equipop}}. By sampling the space of non-overlapping redshift bin configurations we attempt to find better configurations with feasible computation times even while employing a realistic pipeline of analysis.\\

Finding an optimal tomographic configuration for 3x2pt analysis requires taking into consideration source and lens galaxies' mutual dependence. Since the optimal tomographic configuration for the source galaxies depends on the configuration for the lens galaxies and vice-versa, both configuration spaces should be sampled simultaneously. In practice, this is much less efficient, as we discuss in Appendix\,\ref{appendix: simultaneous optimization}, thus an alternative method must be employed. We decided on sampling by alternating between the source and lens bin configurations iteratively. That is, we run our Fisher pipeline of analysis fixing the tomography of the source galaxies while varying the lens' or vice-versa, sampling 200 randomly-perturbed tomographic configurations per iteration. The perturbations to the tomographic configuration consist in a displacement of the edges of each bin in the form of a Gaussian noise with a standard deviation equal to the half-width of the bin, discarding overlapping edges. After running each iteration, the tomographic configuration that yields the highest FoM for a given sample is fixed while perturbed configurations of the other sample are explored, repeating this until the FoM converges.\\

\begin{figure}
  \centering
    \includegraphics[width=\linewidth]{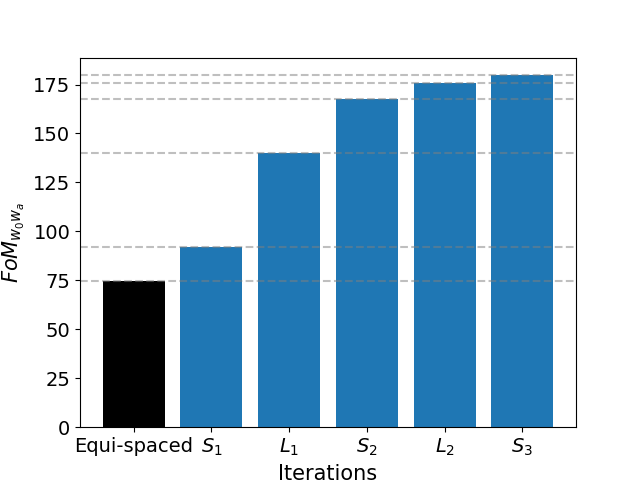}
    \caption{Improvement on the FoM 
    for $w_{\rm 0} w_a$ as we iteratively optimise the source ($S_i$) and lens ($L_i$) tomography.}
 \label{fig:bargraph_8bins}
\end{figure}

In Fig.\,\ref{fig:bargraph_8bins} we show the performance of this method by optimising the tomography for a fiducial case using 8 redshift bins. Our reference is the FoM for equally-spaced redshift bins and the improvement in FoM increases over the first few iterations over source and lens tomographic optimisations before rapidly converging, demonstrating the effectiveness of our method in reaching an optimised redshift bin. The computational resources needed to perform a single iteration are 1200 CPU hours for the most expensive case with 10 redshift bins. \\

The evolution of the optimised FoM over 5 iterations is similar for other number of redshift bins, coinciding in that, after two iterations sampling the source galaxies binning and two iterations sampling lens galaxies binning, the algorithm has basically converged. Based on this, we implement our method for different science cases and different numbers of redshift bins limited to 4 iterations.

\section{Results}\label{results}

Having validated the methodology for the optimisation method and determined a convergence criteria based on a set of test cases, we then proceeded to apply it to different amounts of redshift bins (from 3 to 10) and optimise for two different targets: the FoM for $w_{\rm 0} w_a$ alone, optimising for the improvement of the constraints for the two parameters irrespective of the effect on the constraints on the rest of the parameters of the model, and the FoM for the whole set of cosmological parameters of $w_{\rm 0} w_a\text{CDM}$.

\subsection{Tomography optimisation for 3x2pt analysis: $w_{\rm 0} w_a$}

Differentiating evolving DE models from $\Lambda\text{CDM}$ requires a precise measurement of the dynamics of structure formation over very large time scales. For this reason, low-redshift surveys struggle to detect such subtle effects. Stage-IV surveys that reach higher redshifts will be able to better characterise the $w_{\rm 0}$ and $w_a$ parameters. We show that optimising the tomographic configuration specifically to maximize information about the DE equation of state parameters results in substantial improvements in their constraints. \\

\begin{figure}
  \centering
  \begin{subfigure}{.5\textwidth}
    \centering
    \includegraphics[width=0.95\linewidth]{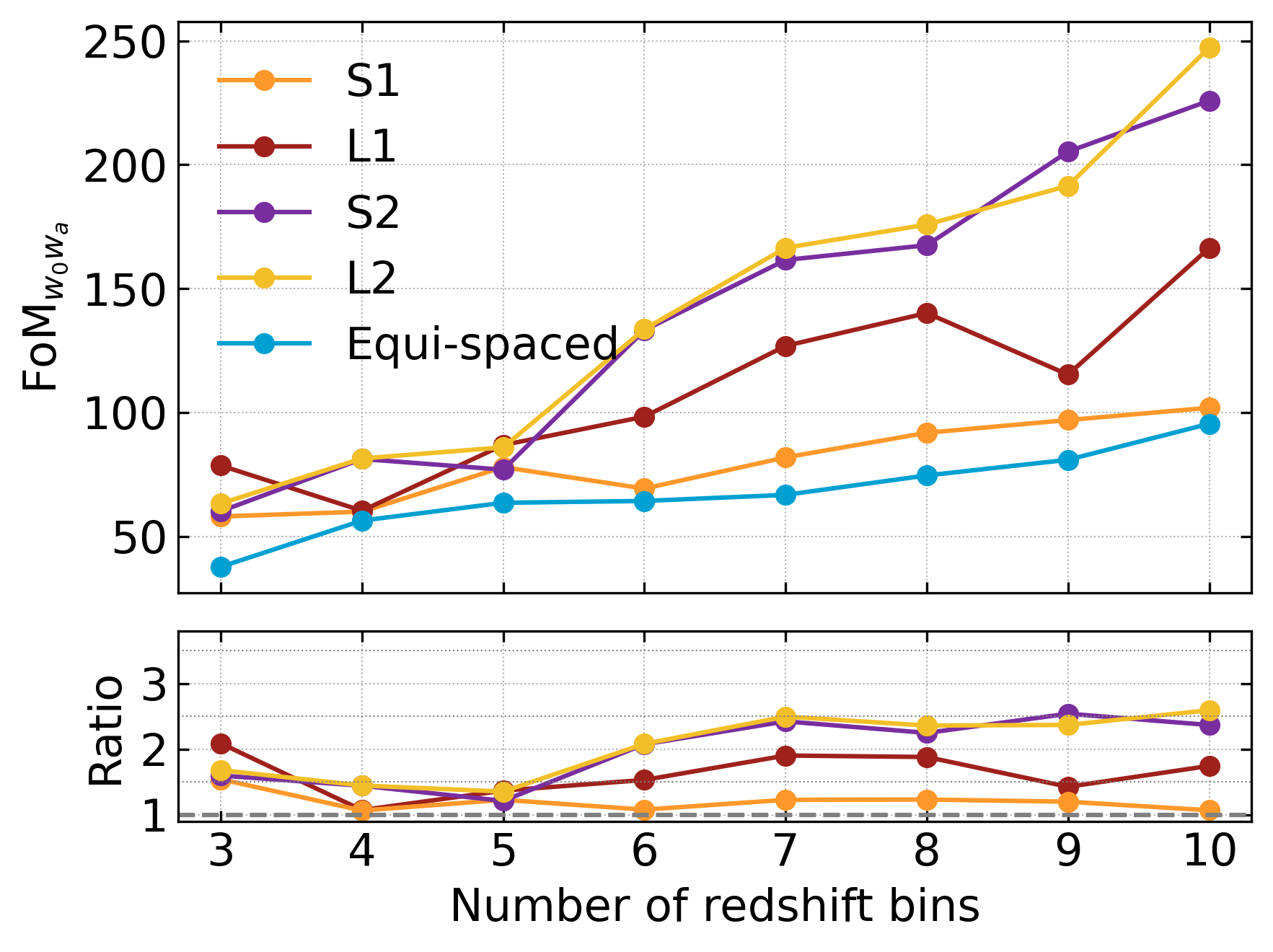}  
    \caption{$\text{FoM}_{w_{\rm 0}\,w_a}$ optimisation for four iterations of the optimisation. The metric in this case quantifies information just for $w_{\rm 0}$ and $w_a$.}
    \label{fig:3x2pt_optimization_w0wa}
  \end{subfigure}
  \hfill
  \begin{subfigure}{.5\textwidth}
    \centering
    \includegraphics[width=0.95\linewidth]{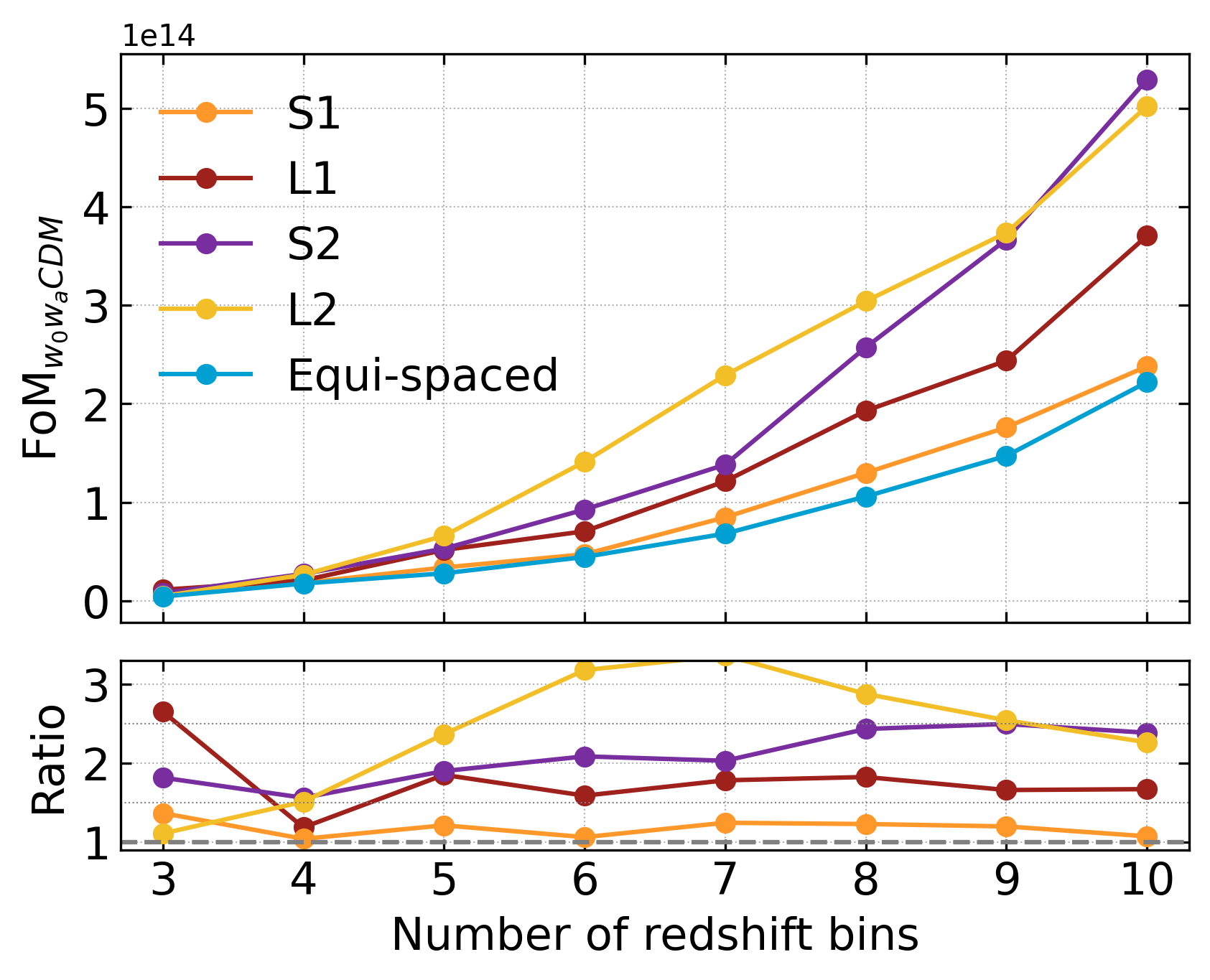}  
    \caption{$\text{FoM}_{w_{\rm 0}\,w_a}$ optimisation for four iterations of the optimisation. The metric in this case quantifies information for the seven cosmological parameters of $w_{\rm 0}w_a\text{CDM}$. }
    \label{fig:3x2pt_optimization_w0waCDM}
  \end{subfigure}
  \caption{Improvement on information as the tomography is iteratively optimised.}
  \label{convergence_plots}
\end{figure}

The first iteration of the optimisation process samples different configurations for the tomography of the source galaxies while fixing the tomographic configuration of the lens galaxies to the equally-spaced case. As we see in the "S1" line in Fig.\,\ref{fig:3x2pt_optimization_w0wa} this first sampling of only the source sample does not lead to a major improvement in the FoM. However, with the second iteration of the algorithm, fixing the source sample to the best-performing tomography and sampling the lens configurations, the improvement in our metric increases further. Subsequent iterations may help with the stability of the optimisation by reducing the variance in the FoM of the optimised configuration but do not increase the FoM any further.\\

The resulting FoM after applying our optimisation method is shown in Fig.\,\ref{fig:3x2pt_optimization_w0wa}, where we see an improvement in our FoM of around 2-fold for configurations including a relatively large number of redshift bins ($N_{bins}>5$). A certain degree of variability in the convergence is present, mostly at a lower number of redshift bins. Despite the inherent variation in a stochastic optimisation algorithm, the improvement is consistent across different numbers of iterations and redshift bins. This result illustrates the importance of selecting an optimal tomographic configuration in modern cosmological probes, as the typical case of equally-spaced bins is significantly outperformed by our optimal configuration of the source and lens redshift bins. In particular, we find $\sim 25\%$ improvement on the individual constraints for both $w_{\rm 0}$ and $w_a$. Similar results are obtained when simultaneously varying the non-DE model parameters, see Fig.\,\ref{fig:3x2pt_optimization_w0waCDM}\\

\begin{figure*}
    \centering
    \begin{subfigure}{0.45\textwidth}
        \centering
        \includegraphics[width=\textwidth]{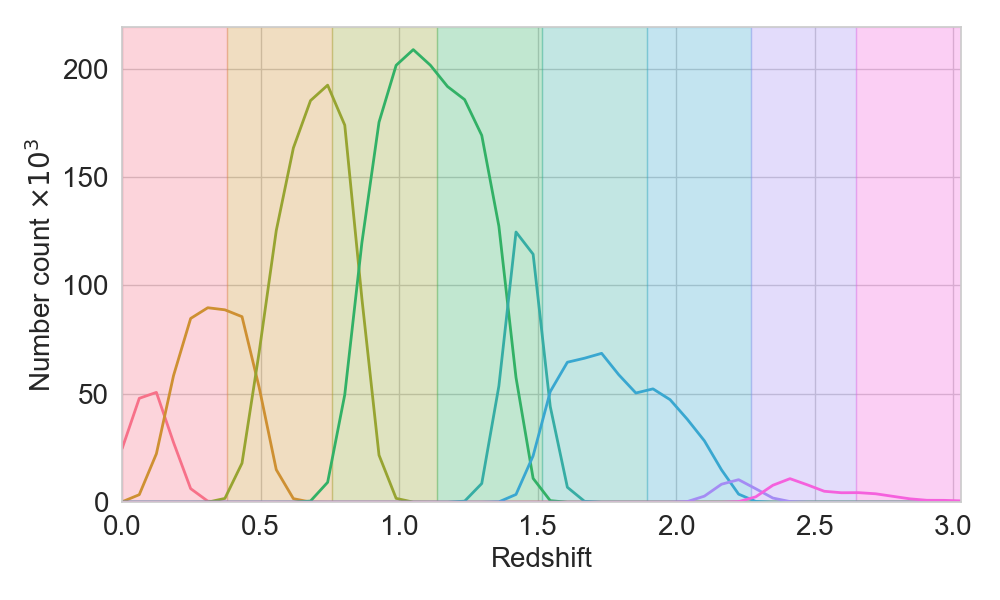}
        \caption{Galaxy number counts for the source sample after a first iteration (S1).}
        \label{fig:subfig1}
    \end{subfigure}
    \hfill
    \begin{subfigure}{0.45\textwidth}
        \centering
        \includegraphics[width=\textwidth]{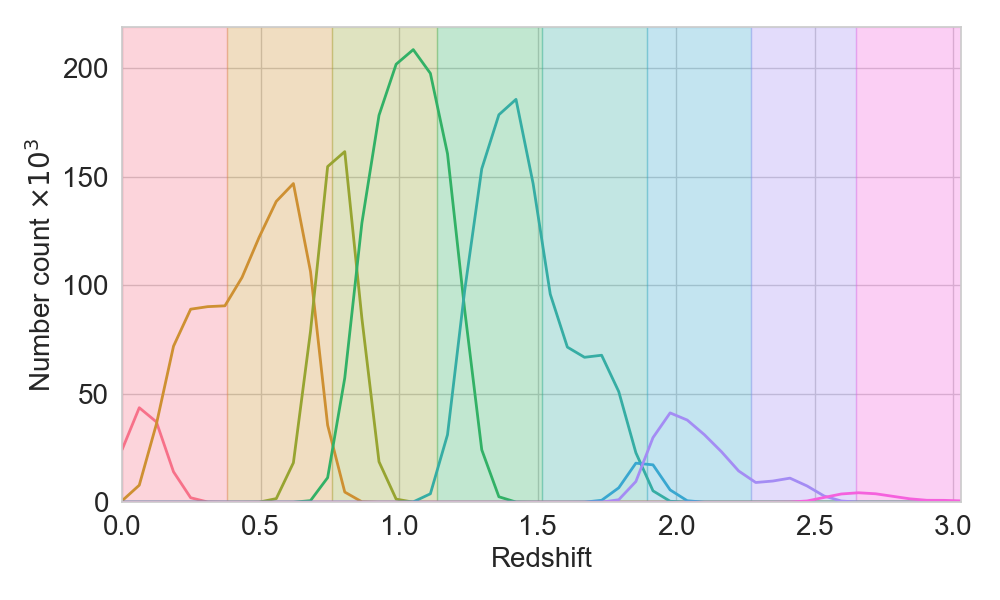}
        \caption{Galaxy number counts for the lens sample after a first iteration (L1).}
        \label{fig:subfig2}
    \end{subfigure}
    
    \vspace{0.4cm}
    
    \begin{subfigure}{0.45\textwidth}
        \centering
        \includegraphics[width=\textwidth]{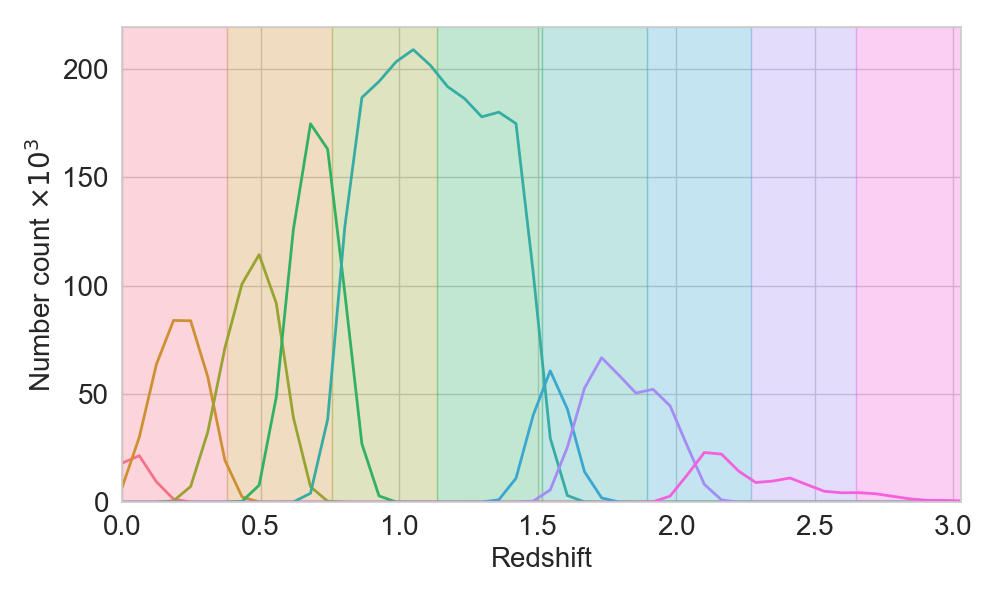}
        \caption{Galaxy number counts for the source sample after a second iteration (S2).}
        \label{fig:subfig3}
    \end{subfigure}
    \hfill
    \begin{subfigure}{0.45\textwidth}
        \centering
        \includegraphics[width=\textwidth]{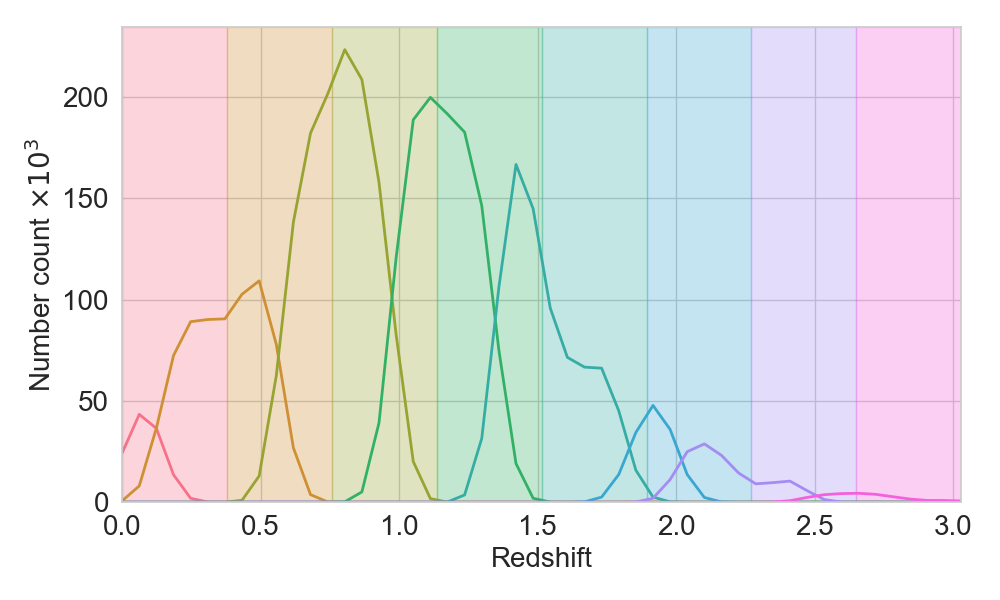}
        \caption{Galaxy number counts for the lens sample after a second iteration (L2).}
        \label{fig:subfig4}
    \end{subfigure}
    
    \caption{Galaxy number counts for source (left) and lens (right) galaxies for 8 redshift bins over several iterations of the optimisation method applied for the dynamical DE model parameters $w_{\rm 0} w_a$. The coloured bands represent the bin edges for the base case of equally-spaced bins.}
    \label{fig:nofzs_4iter_w0wa}
\end{figure*}

As for insights into the physical interpretation of the improvement in the constraints for $w_{\rm 0} w_a$, we have to look at the optimised tomographic configurations as seen in Fig.\,\ref{fig:nofzs_4iter_w0wa}. In Fig.\, \ref{fig:subfig1} and Fig.\, \ref{fig:subfig3}, we show the impact of changing the source redshift bins. The optimal configuration converges towards wider redshift bins at high redshift. In contrast, for the lens samples, the optimisation yields thinner lens bins at the low-to-mid redshift range, as displayed in Fig.\,\ref{fig:subfig2} and Fig.\,\ref{fig:subfig4}. This is understandable: lens bin widths will tend to adjust to the photometric error, $\sigma_{photo-z}$, to maximize the signal in clustering measurements \citep{photoz_galaxyclustering}. In contrast, obtaining a higher signal in WL implies larger bins at high redshift, where fewer galaxies are present, and where the photo-$z$ estimations will be worse but less impactful \citep{wl_degeneracies}.\\

Since our photo-$z$ catalogue contains very low-redshift galaxies where the WL signal will be lower, the tendency of the algorithm is to disregard those galaxies into a very low-redshift bin with little contribution to the overall constraining power. This is best seen in Fig.\,\ref{fig:subfig3} in the optimisation step involving the second source bin iteration. In the case of Fig.\,\ref{fig:subfig4}, corresponding to the second iteration of the lens sample, the low-redshift bin is suppressed by the optimisation algorithm, favouring instead a bigger number density in the mid-range redshift region lens bins. It is also worth noting the tendency to create a wide bin of source galaxies at the redshift mid-range. Given that the lensing signal is larger with lens galaxies halfway (in terms of redshift) to the source galaxies, this suggests that an optimal tomography for 3x2pt favours a larger lensing signal at high and mid redshifts, even at the cost of worsening the clustering for wider lens bins. Although these tendencies are hinted in Fig. \ref{fig:subfig1} and Fig. \ref{fig:subfig2}, it is only when further iterations are implemented that this effect is fully realised (see Figs.\,\ref{fig:subfig3} and \ref{fig:subfig4}. In fact, the complex interplay between the optimisation of lensing and clustering signals at different redshift ranges is precisely the reason why an efficient numerical tomographic optimisation method needs to be devised.\\

The initial assessment of 4 iterations being sufficient to converge into an optimised tomography is compatible with the results for different number of redshift bins. Using a larger number of tomographic bins, i.e, ranging from 6 to 10 bins, in Fig. \ref{fig:3x2pt_errorbars_a} we find that there are significant changes with respect to the fiducial tomographic bin configuration, that is, the changes in FoM are larger than the estimated variance of the optimisation. We also find that by the fourth iteration (L2) the gains have stabilised. For this reason we consider the optimisation method to converge after 4 iterations. \\

\subsection{Tomography optimisation for 3x2pt analysis: $w_{\rm 0} w_a \text{CDM}$}
\label{sec:tomopt}

In the case where we optimise for the whole set of parameters of our cosmological model, that is $w_{\rm 0}$, $w_a$, $\Omega_{{\rm m}}$, $\Omega_{\rm b}$, $n_{\rm s}$, $\text{ln}(10^{10}\text{A}_{\rm s})$ and $h$, we expect a priori a higher improvement in the FoM for an optimised tomography with respect to the base case of equally-spaced bins by virtue of the higher number of parameters. However, by looking at Fig.\,\ref{fig:3x2pt_optimization_w0waCDM} the relative gain compared with the optimisation of $w_{\rm 0} w_a$ is only modestly better hovering at around 2.5-fold. In reality there is an interplay where the requirements to constrain certain parameters which are more sensitive to low-redshift growth, such as $\sigma_8$ (which we sample indirectly through $\text{A}_{\rm s}$) or $\Omega_{\rm m}$, and the requirements to constrain those parameters more sensitive to high redshift evolution, such as $w_a$ may differ in terms of optimal tomographic binning.\\

\begin{figure}
  \centering
  \begin{subfigure}{.5\textwidth}
    \centering
    \includegraphics[width=0.95\linewidth]{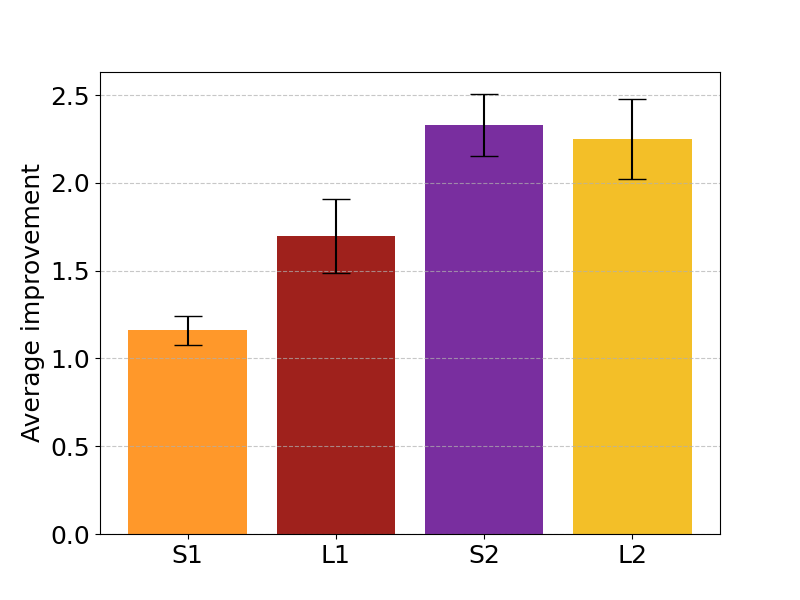}  
    \caption{Average improvement and standard deviation from 6 to 10 redshift bins over each iteration of the optimisation method targeting the dynamical DE parameters $w_{\rm 0} w_a$.}
    \label{fig:3x2pt_errorbars_a}
  \end{subfigure}
  \hfill
  \begin{subfigure}{.5\textwidth}
    \centering
    \includegraphics[width=0.95\linewidth]{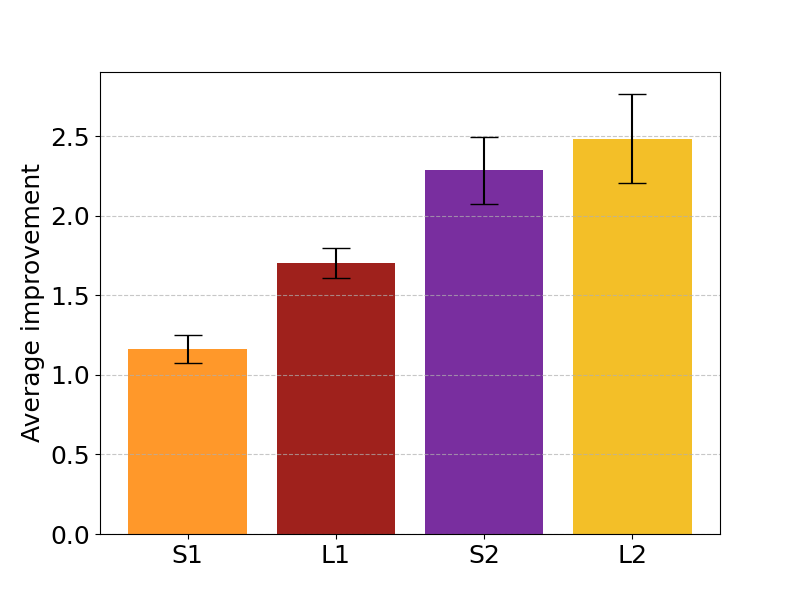}  
    \caption{Average improvement and standard deviation from 6 to 10 redshift bins over each iteration of the optimisation method targeting the full set of cosmological parameters of $w_{\rm 0} w_a \text{CDM}$.}
    \label{fig:3x2pt_errorbars_b}
  \end{subfigure}
  \caption{Average improvement over iterations giving a rough estimate of the variance introduced by the stochasticity of the optimisation algorithm.}
  \label{fig:3x2pt_errorbars}
\end{figure}

Given that the number of parameters and hence the dimensionality of the parameter space increases, the improvement over iterations as well as the absolute gain in FoM after the method is applied have less variability. This smoother convergence over iterations is expected due to the dependence of the FoM over a larger set of parameters that includes parameters less sensitive than $w_{\rm 0}$ and $w_a$ to changes in the tomography. Despite that, we show in Fig.  \ref{fig:3x2pt_errorbars_b} that the average improvement over a different number of redshift bins is similar to the case when we optimise for $\text{FoM}_{w_{\rm 0} w_a}$. This reiterates the consistency of the algorithm for different target parameters.

\subsection{Parameter optimisation dependence}

\begin{figure}
  \centering
  \begin{subfigure}{.5\textwidth}
    \centering
    \includegraphics[width=0.95\linewidth]{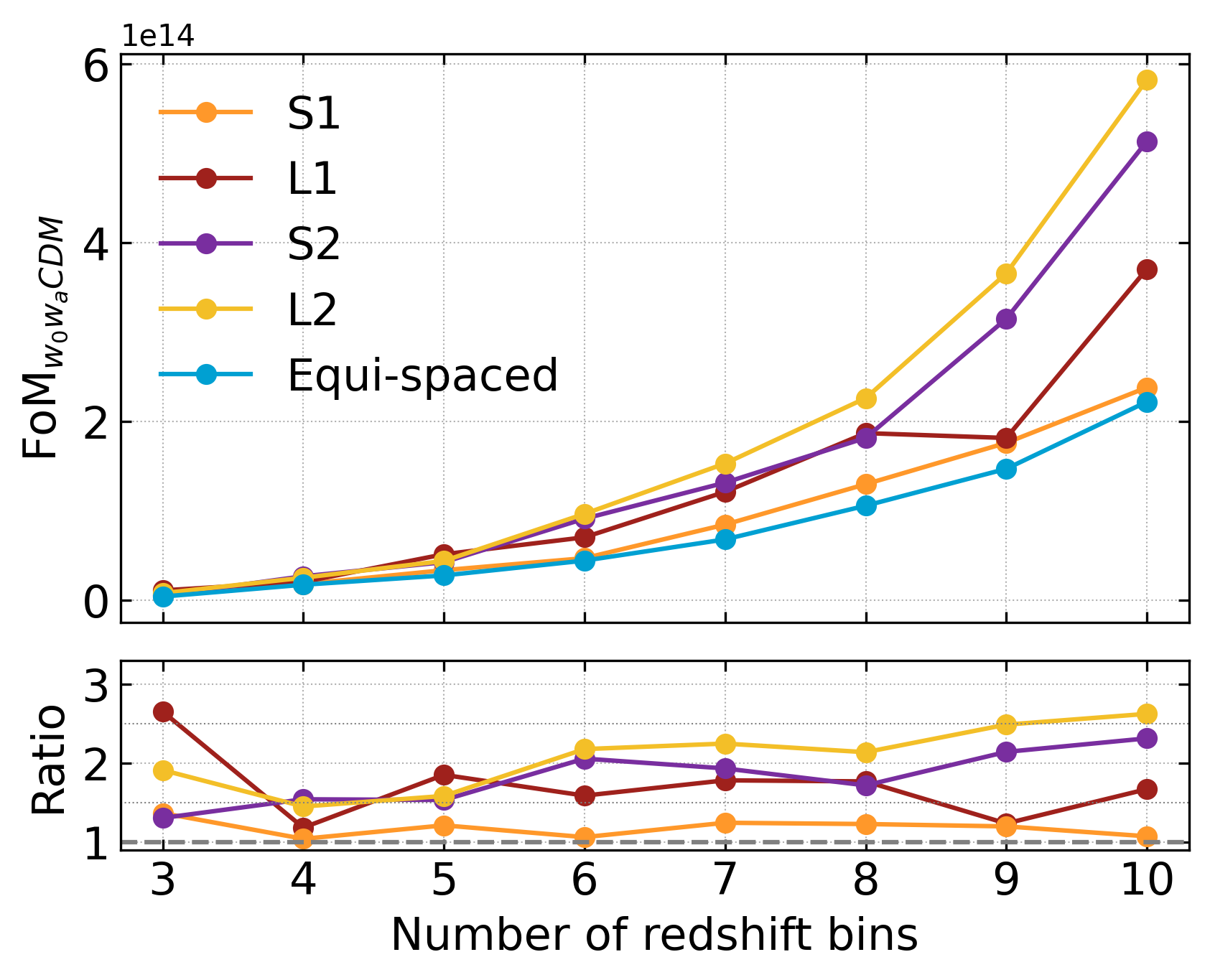}  
    \caption{Computing the $\text{FoM}_{w_{\rm 0}\,w_a \text{CDM}}$ metric for the optimisation targeting $w_{\rm 0}w_a$.}
    \label{fig:3x2pt_target_dependence_w0wacosmo}
  \end{subfigure}
  \hfill
  \begin{subfigure}{.5\textwidth}
    \centering
    \includegraphics[width=0.95\linewidth]{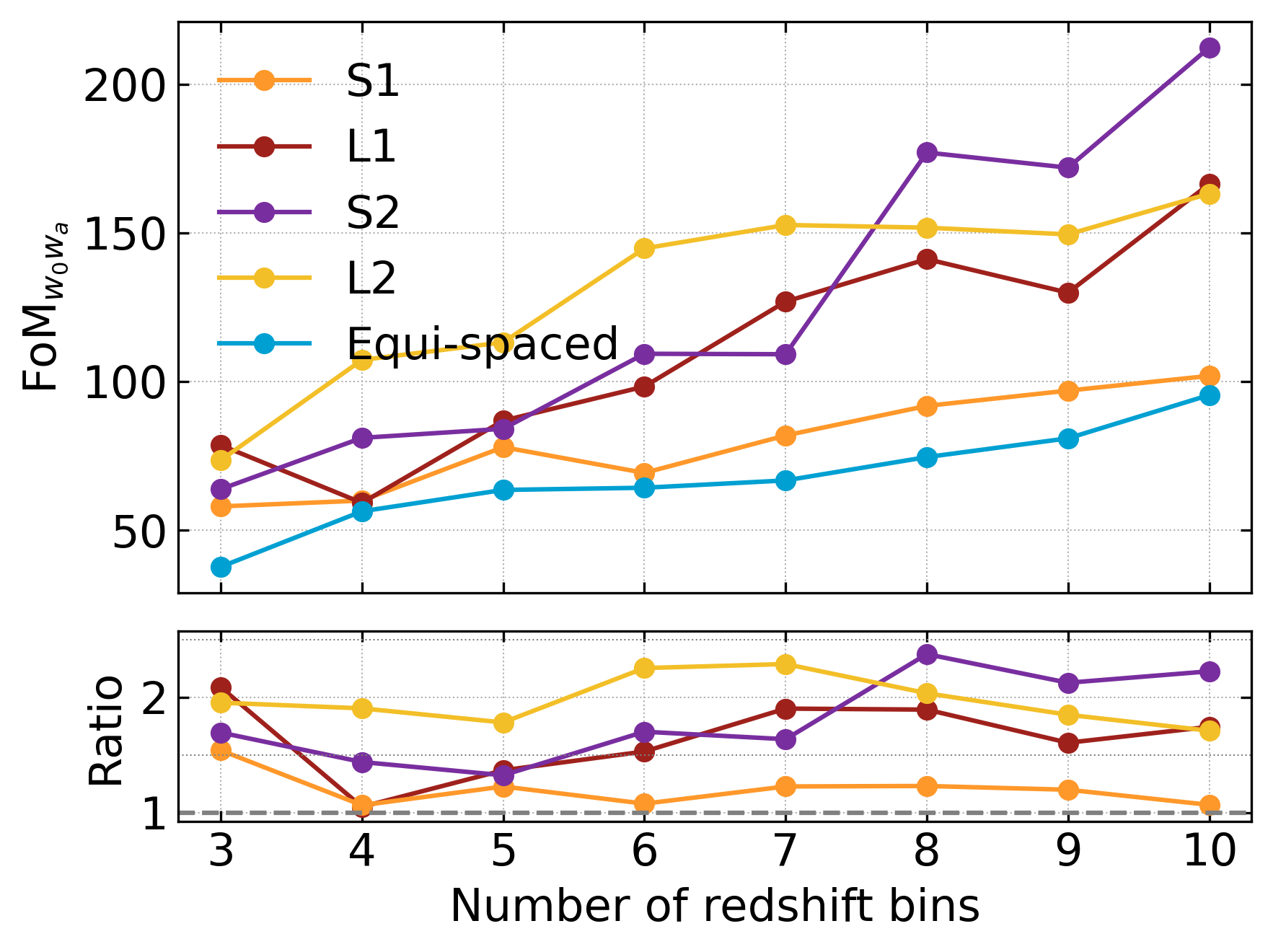}  
    \caption{Computing the $\text{FoM}_{w_{\rm 0}\,w_a}$ metric for the optimisation targeting $w_{\rm 0}w_a \text{CDM}$.}
    \label{fig:3x2pt_target_dependence_cosmow0wa}
  \end{subfigure}
  \caption{Improvement of off-target cosmological parameter constrains after applying our optimisation method.}
  \label{fig:3x2pt_target_dependence}
\end{figure}

Since our optimisation target, the FoM, depends on which parameters are marginalised, the impact of the optimisation on the parameters that are marginalised over must be assessed. To do that we have computed the FoMs with different marginalised parameters than those used during the optimisation. That is, having used the two metrics FoM$_{w_{\rm 0}w_a}$ and FoM$_{w_{\rm 0}w_a \text{CDM}}$ marginalising over all but two or all but seven cosmological parameters, respectively, it could be expected that the optimisation for $w_{\rm 0} w_a$ could be at the cost of losing constraining power over other cosmological parameters. In order to assess that, we compute the $\text{FoM}_{w_{\rm 0}w_a \text{CDM}}$ for the Fisher matrices obtained when optimising for $\text{FoM}_{w_{\rm 0}w_a}$ and vice-versa. The results shown in Fig.\,\ref{fig:3x2pt_target_dependence} suggest, on the one hand, that using the target of $w_{\rm 0} w_a$ in our optimisation does not result in a significant decrease of constraining power for the combined set of cosmological parameters. On the other hand, when targeting the whole set of $w_{\rm 0} w_a \text{CDM}$ parameters in our optimisation, most of the improvement in constraints must lie in the constraints for $w_{\rm 0} w_a$.\\

We explored the effect of the optimisation for a given set of parameters (on-target) on the constraints for the rest of the parameters not included in the FoM metric (off-target). In order to do that we measured the FoM for the on-target and off-target parameters after the optimisation process and we compared the relative improvement with respect to the case with equally-spaced binning. In Table \ref{table:target dependence} we show the relative improvement for the on-target and off-target FoM. The expectation is that the improvement should be highest for the on-target parameters, although the stochastic nature of the method introduces some variation in the relative improvement. For example, performing the optimisation targeting $(w_{\rm 0}, w_a)$ results in a relative improvement of $\Delta \text{FoM}_{w_{\rm 0}w_a} =2.45$, larger than the off-target improvement computing the metric for the set of parameters of $w_{\rm 0}w_a\text{CDM}$, with $\Delta \text{FoM}_{w_{\rm 0}w_a\text{CDM}} =2.17$. This similar albeit smaller relative improvement of off-target parameters indicates that the optimisation method does not result in worsening the constraints of the off-target parameters of the model. We discuss in more detail the impact of the optimisation on off-target cosmological parameters in Sect. \ref{appendix: likelihoods} in the appendix. We also present the case where the single probe of WL was used in the analysis instead of 3x2pt. In this case, the constraining power is smaller to begin with, resulting in much greater variance and a case where the off-target improvement of $\Delta \text{FoM}_{w_{\rm 0}w_a\text{CDM}} =4.78$ is larger than on-target. From this we conclude that our method performs most consistently when the constraining power is larger.\\

\begin{table*}
\centering
\caption{Comparison of on-target and off-target improvement in our metric after applying our optimisation method. The improvement with respect to the base case of equally-spaced bins is given by $\Delta \text{FoM}$.}
\label{table:target dependence}

\begin{tabular}{llcc}
\toprule
\textbf{Analysis} & \textbf{Optimisation target} & $\Delta \text{FoM}_{w_{\rm 0} w_a}$ & $\Delta \text{FoM}_{w_{\rm 0} w_a \text{CDM}}$ \\ \cline{1-4}
\midrule
\multirow{2}{*}{3x2pt (8 redshift bins)} 
    & $w_{\rm 0} w_a$              & 2.45 & 2.17 \\
    & $w_{\rm 0} w_a \text{CDM}$          & 2.22 & 2.25 \\
\midrule \cline{1-4}
\multirow{2}{*}{WL only (8 redshift bins)} 
    & $w_{\rm 0} w_a$              & 3.16 & 4.78 \\
    & $w_{\rm 0} w_a \text{CDM}$          & 2.49 & 3.90 \\
\bottomrule
\end{tabular}
\end{table*}

\section{Discussion}\label{discussion}

The characterisation of DE is one of the most ambitious goals of current physics. A significant amount of resources has been invested for that objective. For example, recent DESI results \citep{DESI_Y2} have hinted at possible deviations from $\Lambda \text{CDM}$ from the combination of measurements of baryon acoustic oscillations (BAO), type-Ia supernovae measurements, and cosmic microwave background (CMB) data. Their results favoured $w_{\rm 0}w_a\text{CDM}$ with a significance of $2.8-4.2 \sigma$. While DESI used spectroscopic data to study LSS, other surveys will use a complementary approach, based on photometric data to test dynamical DE models. However, optimising the tomographic sample selection of photometric galaxy surveys for the study of DE is still an open question.\\

Our work provides a rather general end-to-end cosmological inference pipeline to optimise dark-energy constraints from 3x2pt analyses for next generation galaxy surveys. In particular, we explore configurations beyond the fiducial paradigm of equal-width or equally-populated redshift bins in a fast and robust way. In doing so, we show that photometric surveys could benefit from incorporating optimised tomographic selections in their analysis procedure, at least doubling the DE FoM thanks to our optimisation method.\\

We have implemented our pipeline for a wide range of redshift bin configurations. The reason is twofold: to show the robustness of the method over many analysis setups and to assess the impact of an optimisation as the number of tomographic bins is increased. We have found that our method is indeed robust and our convergence criteria sufficient for six or more redshift bins. The degree of relative improvement of the FoM after the four iterations was similar for six or more redshift bins as well.\\

We performed our optimisation method using our own stage-IV-like photometric catalogue and with a fairly realistic analysis pipeline, with results consistent with \cite{Euclid_blanchard} for the same number of redshift bins. Thus, we expect our improvement to be applicable in future analysis, implying an improvement in the constraints of the individual parameters of $w_{\rm 0}$ and $w_a$ of about 25\%.\\

Previous work from \cite{andreas_paper} compared equally-spaced tomographic bins and equally-populated bins finding an optimal configuration for each number of redshift bins. They found that, as the number of redshift bins changes, the optimal case between equally- populated and equally-spaced bin configurations can change as well, similarly to what we found (see Fig. \ref{fig:fiducial}). However, the relative improvement between the two configurations in their case was smaller than with our method. The work of \cite{Taylor:2018nrc} also explored the issue of using the best tomographic binning strategy. The authors noted, in their particular case using a larger number (twenty) of redshift bins, that equally-populated redshift bins can lead to excessively narrow bins in the region of the $n(z)$ with most galaxies, while equally-spaced bins result in a loss of information due to a low number of galaxies at high redshift. While they opted for equally-spaced bins as optimal in their setup, their work again highlights the need to be flexible in the way we approach the issue of choosing our tomographic redshift bins.\\

In the work by \cite{wong_3x2pt_optimization}, an exploration of the optimal tomographic configuration of 3x2pt analysis for \textit{Euclid}, an ongoing stage-IV survey, was performed. The authors performed their exploration with equally-spaced, equally-populated, and equal comoving distance redshift bins. They used a realistic setup using mock realisations of \textit{Euclid} observations and concluded that the best-performing tomography for 3x2pt analysis was equally-populated bins, with information saturating in $\ge 7-8$ redshift bins. Information saturation in a lower number of redshift bins would be expected with the settings of the first \textit{Euclid} data release, with a smaller sky coverage of $2600$ square degrees. In contrast, our work explores a range of tomographic configurations beyond those three. We explored tomographic binning for the source and lens samples separately and, most importantly, in an iterative manner that allows for a quick convergence. We also ensured that the choice of initial conditions does not have an impact on the convergence, as discussed Sect.\,\ref{sec:equipop} in the appendix. Our results show that we can improve the amount of information on $w_{\rm 0}$ and $w_a$ by using the highly customised tomographic bins that we obtained. While our method to define tomographic configurations to explore is numerical, in Sect. \ref{sec:tomopt} we present possible physical insights for the optimal configurations of both the source and lens samples. \\

The CosmoDC2 simulation was also used in the LSST-DESC 3x2pt Tomography Optimization Challenge \citep{LSST_3x2pt_optimization}, where a variety of tomographic configuration algorithms were compared in terms of their ability to improve the information on $w_{\rm 0}$ and $w_a$ using 3x2pt analysis. Of the methods presented there, some optimised the assignment of galaxies to fixed tomographic bins and others optimised the configuration of bin edges. We compare our method with the latter. Highlighting the complexity of the problem, the optimised edges for 3x2pt failed to yield good results for lensing alone. This points to the need of optimising the tomography for specific science cases. However, the methods that used fixed bins and optimised the classification of galaxies into bins plateaued at a metric value of 120-140, inferior to the best-performing methods that optimised the bin edges and reached up to 167. This result strongly emphasizes the need to explore redshift bin selections beyond the usual equally spaced or equally populated choices. In this work we used a different setup using more photometric bands, different footprints, and a much smaller training set ($\approx10^5$ galaxies to train our SOM vs. $\approx10^6$ in their training), but we obtained a larger relative improvement with our optimised redshift bin selection with respect to the equally-spaced tomography.\\

We emphasize that, although the method presented here is a proof of concept, it already yields substantial improvements in the constraints for the cosmological parameters in $\Lambda \text{CDM}$ and dynamical DE cosmologies. Increasing the FoM for $w_{\rm 0} w_a$ by a factor 2 or more would be comparable to a substantial increase in the sky coverage of a survey. Even acknowledging that forecasts in the Fisher formalism have potential limitations, the improvement that we can expect from applying our method to real data and a full Markov chain Monte-Carlo analysis should be of interest for any current and future photometric surveys aiming at providing competitive cosmological constraints.

\section{Conclusions}\label{conclusion}

We have developed a new method to optimise the way in which the tomographic redshift bins for 3x2pt analyses in photometric galaxy surveys are selected. As a proof of concept, we have generated a photo-$z$ galaxy catalogue from the CosmoDC2 simulation using a SOM and we have selected the training and photo-$z$ galaxy catalogues with similar characteristics to current or planned stage-IV surveys. We have also used a pipeline of cosmological parameter inference to estimate the contours for a set of cosmological DE-related parameters using the Fisher formalism. We have used the FoM metric derived from the marginalized Fisher matrix to assess the constraining power of a given choice of tomographic redshift bins in our analysis, and we have optimised the sample selection for both the source and the lens galaxy samples separately and in an iterative manner. We have done so using sets of 200 perturbed bin edges per iteration and have found that the method consistently converges after four iterations, two of the source sample and two of the lens sample. We have selected the $w_{\rm 0} w_a$ parameters of dynamical DE as our target for optimisation, but we have found that the improvement in the constraints of those parameters does not come at the cost of a decreased constraining power for the rest of the parameters of the model. In contrast, there is an additional improvement in other cosmological parameter's constraints. Overall, we have found an average improvement for $w_{\rm 0}$ and $w_a$ of $\approx 25$\% based on the FoM values, although Fisher-derived contours suggest that the improvement is most prominent for $w_a$ in particular. Furthermore, we have shown the robustness of our optimisation method against different analysis choices. In conclusion, we propose the sample selection optimisation pipeline presented here as a powerful tool to optimise the scientific return of next-generation photometric galaxy surveys.

\begin{acknowledgements}
PF and MA acknowledge support form the Spanish Ministerio de Ciencia, Innovaci\'on y Universidades, projects PID2019-11317GB, PID2022-141079NB, PID2022-138896NB; the European Research Executive Agency HORIZON-MSCA-2021-SE-01 Research and Innovation programme under the Marie Skłodowska-Curie grant agreement number 101086388 (LACEGAL) and the programme Unidad de Excelencia Mar\'{\i}a de Maeztu, project CEX2020-001058-M.
This work has made use of CosmoHub, developed by PIC (maintained by IFAE and CIEMAT) in collaboration with ICE-CSIC. It received funding from the Spanish government (grant EQC2021-007479-P funded by MCIN/AEI/10.13039/501100011033), the EU NextGeneration/PRTR (PRTR-C17.I1), and the Generalitat de Catalunya. Ayuda PRE2020-094899  de la ayuda financiada por MCIN/AEI/10.13039/501100011033 y por FSE invierte en tu futuro.
\end{acknowledgements}

\bibliographystyle{mnras}
\bibliography{example}

\begin{thebibliography}{}
\makeatletter
\relax
\def\mn@urlcharsother{\let\do\@makeother \do\$\do\&\do\#\do\^\do\_\do\%\do\~}
\def\mn@doi{\begingroup\mn@urlcharsother \@ifnextchar [ {\mn@doi@} {\mn@doi@[]}}
\def\mn@doi@[#1]#2{\def\@tempa{#1}\ifx\@tempa\@empty \href {http://dx.doi.org/#2} {doi:#2}\else \href {http://dx.doi.org/#2} {#1}\fi \endgroup}
\def\mn@eprint#1#2{\mn@eprint@#1:#2::\@nil}
\def\mn@eprint@arXiv#1{\href {http://arxiv.org/abs/#1} {{\tt arXiv:#1}}}
\def\mn@eprint@dblp#1{\href {http://dblp.uni-trier.de/rec/bibtex/#1.xml} {dblp:#1}}
\def\mn@eprint@#1:#2:#3:#4\@nil{\def\@tempa {#1}\def\@tempb {#2}\def\@tempc {#3}\ifx \@tempc \@empty \let \@tempc \@tempb \let \@tempb \@tempa \fi \ifx \@tempb \@empty \def\@tempb {arXiv}\fi \@ifundefined {mn@eprint@\@tempb}{\@tempb:\@tempc}{\expandafter \expandafter \csname mn@eprint@\@tempb\endcsname \expandafter{\@tempc}}}

\bibitem[\protect\citeauthoryear{Abbott, Aguena, Alarcon  et~al.}{Abbott et~al.}{2022}]{DESY3}
Abbott T. M.~C.,  Aguena M.,  Alarcon A.,   et~al., 2022, \mn@doi [PRD] {10.1103/PhysRevD.105.023520}, 105, 023520

\bibitem[\protect\citeauthoryear{Abdul~Karim, Aguilar, Ahlen  et~al.}{Abdul~Karim et~al.}{2025}]{DESI_Y2}
Abdul~Karim M.,  Aguilar J.,  Ahlen S.,   et~al., 2025, arXiv:2503.14738

\bibitem[\protect\citeauthoryear{Abolfathi, Alonso, Armstrong  et~al.}{Abolfathi et~al.}{2021}]{DESCDC2}
Abolfathi B.,  Alonso D.,  Armstrong R.,   et~al., 2021, \mn@doi [ApJS] {10.3847/1538-4365/abd62c}, 253, 31

\bibitem[\protect\citeauthoryear{Albrecht, Bernstein, Cahn  et~al.}{Albrecht et~al.}{2006}]{fom_ref}
Albrecht A.,  Bernstein G.,  Cahn R.,   et~al., 2006, arXiv:astro-ph/0609591

\bibitem[\protect\citeauthoryear{Asgari, Lin, Joachimi  et~al.}{Asgari et~al.}{2021}]{KiDS:2020suj}
Asgari M.,  Lin C.-A.,  Joachimi B.,   et~al., 2021, \mn@doi [A\&A] {10.1051/0004-6361/202039070}, 645, A104

\bibitem[\protect\citeauthoryear{Blazek, MacCrann  \& Troxel}{Blazek et~al.}{2019}]{Blazek:2017wbz}
Blazek J.,  MacCrann N.,   Troxel M.~A.,  2019, \mn@doi [PRD] {10.1103/PhysRevD.100.103506}, 100, 103506

\bibitem[\protect\citeauthoryear{{Carretero} et~al.,}{{Carretero} et~al.}{2017}]{2017ehep.confE.488C}
{Carretero} J.,  et~al., 2017, in Proceedings of the European Physical Society Conference on High Energy Physics. 5-12 July. p.~488, \mn@doi{10.22323/1.314.0488}

\bibitem[\protect\citeauthoryear{Chevallier \& Polarski}{Chevallier \& Polarski}{2001}]{dynamical_dark_energy_2}
Chevallier M.,  Polarski D.,  2001, \mn@doi [Int. J. Mod. Phys. D] {10.1142/S0218271801000822}, 10, 213

\bibitem[\protect\citeauthoryear{Dalal, Li, Nicola  et~al.}{Dalal et~al.}{2023}]{Dalal:2023olq}
Dalal R.,  Li X.,  Nicola A.,   et~al., 2023, \mn@doi [PRD] {10.1103/PhysRevD.108.123519}, 108, 123519

\bibitem[\protect\citeauthoryear{Desjacques, Jeong  \& Schmidt}{Desjacques et~al.}{2018}]{galaxy_bias}
Desjacques V.,  Jeong D.,   Schmidt F.,  2018, \mn@doi [Phys. Rept.] {10.1016/j.physrep.2017.12.002}, 733, 1

\bibitem[\protect\citeauthoryear{Euclid Collaboration:~Blanchard, Camera, Carbone  et~al.}{Euclid Collaboration:~Blanchard et~al.}{2020}]{Euclid_blanchard}
Euclid Collaboration:~Blanchard A.,  Camera S.,  Carbone C.,   et~al., 2020, \mn@doi [A\&A] {10.1051/0004-6361/202038071}, 642, A191

\bibitem[\protect\citeauthoryear{Euclid Collaboration:~Lepori, Tutusaus, Viglione  et~al.}{Euclid Collaboration:~Lepori et~al.}{2022}]{Euclid:2021rez}
Euclid Collaboration:~Lepori F.,  Tutusaus I.,  Viglione C.,   et~al., 2022, \mn@doi [A\&A] {10.1051/0004-6361/202142419}, 662, A93

\bibitem[\protect\citeauthoryear{Euclid Collaboration:~Mellier, Abdurro'uf{,}, Acevedo~Barroso  et~al.}{Euclid Collaboration:~Mellier et~al.}{2025}]{Euclid:2024yrr}
Euclid Collaboration:~Mellier Y.,  Abdurro'uf{,} Acevedo~Barroso J.,   et~al., 2025, \mn@doi [A\&A] {10.1051/0004-6361/202450810}, 697, A1

\bibitem[\protect\citeauthoryear{Euclid Collaboration:~Pocino, Tutusaus, Castander  et~al.}{Euclid Collaboration:~Pocino et~al.}{2021}]{andreas_paper}
Euclid Collaboration:~Pocino A.,  Tutusaus I.,  Castander F.~J.,   et~al., 2021, \mn@doi [A\&A] {10.1051/0004-6361/202141061}, 655, A44

\bibitem[\protect\citeauthoryear{Fang, Krause, Eifler  et~al.}{Fang et~al.}{2020}]{Fang:2019xat}
Fang X.,  Krause E.,  Eifler T.,   et~al., 2020, \mn@doi [JCAP] {10.1088/1475-7516/2020/05/010}, 05, 010

\bibitem[\protect\citeauthoryear{Heavens, Refregier  \& Heymans}{Heavens et~al.}{2000}]{Heavens:2000ad}
Heavens A.,  Refregier A.,   Heymans C.,  2000, \mn@doi [MNRAS] {10.1046/j.1365-8711.2000.03907.x}, 319, 649

\bibitem[\protect\citeauthoryear{Heitmann, Finkel, Pope  et~al.}{Heitmann et~al.}{2019}]{cosmodc2_halosim}
Heitmann K.,  Finkel H.,  Pope A.,   et~al., 2019, \mn@doi [ApJ] {10.3847/1538-4365/ab4da1}, 245, 16

\bibitem[\protect\citeauthoryear{Heymans, Tröster, Asgari  et~al.}{Heymans et~al.}{2021}]{KiDS1000}
Heymans C.,  Tröster T.,  Asgari M.,   et~al., 2021, \mn@doi [A\&A] {10.1051/0004-6361/202039063}, 646, A140

\bibitem[\protect\citeauthoryear{{Howlett}, {Lewis}, {Hall}  et~al.}{{Howlett} et~al.}{2012}]{CAMB}
{Howlett} C.,  {Lewis} A.,  {Hall} A.,   et~al., 2012, \mn@doi [JCAP] {10.1088/1475-7516/2012/04/027}, \href {https://ui.adsabs.harvard.edu/abs/2012JCAP...04..027H} {04, 027}

\bibitem[\protect\citeauthoryear{Ivezi{\'c}, Kahn, Anthony~Tyson  et~al.}{Ivezi{\'c} et~al.}{2019}]{2019ApJ...873..111I}
Ivezi{\'c} {\v{Z}}.,  Kahn S.~M.,  Anthony~Tyson J.,   et~al., 2019, \mn@doi [ApJ] {10.3847/1538-4357/ab042c}, 873, 111

\bibitem[\protect\citeauthoryear{Kitching, Taylor, Capak  et~al.}{Kitching et~al.}{2019}]{rainbow}
Kitching T.~D.,  Taylor P.~L.,  Capak P.,   et~al., 2019, \mn@doi [PRD] {10.1103/PhysRevD.99.063536}, 99, 063536

\bibitem[\protect\citeauthoryear{Korytov, Hearin, Kovacs  et~al.}{Korytov et~al.}{2019}]{LSSTDarkEnergyScience:2019hkz}
Korytov D.,  Hearin A.,  Kovacs E.,   et~al., 2019, \mn@doi [ApJS] {10.3847/1538-4365/ab510c}, 245, 26

\bibitem[\protect\citeauthoryear{Lewis, Challinor  \& Lasenby}{Lewis et~al.}{2000}]{CAMB_1}
Lewis A.,  Challinor A.,   Lasenby A.,  2000, \mn@doi [ApJ] {10.1086/309179}, 538, 473

\bibitem[\protect\citeauthoryear{Linder}{Linder}{2003}]{dynamical_dark_energy}
Linder E.~V.,  2003, \mn@doi [PRL] {10.1103/PhysRevLett.90.091301}, 90, 091301

\bibitem[\protect\citeauthoryear{Masters, Capak, Stern  et~al.}{Masters et~al.}{2015}]{Masters:2015asa}
Masters D.,  Capak P.,  Stern D.,   et~al., 2015, \mn@doi [ApJ] {10.1088/0004-637X/813/1/53}, 813, 53

\bibitem[\protect\citeauthoryear{Moskowitz, Gawiser, Bault  et~al.}{Moskowitz et~al.}{2023}]{3x2pt_optimization_2}
Moskowitz I.,  Gawiser E.,  Bault A.,   et~al., 2023, \mn@doi [ApJ] {10.3847/1538-4357/accc88}, 950, 49

\bibitem[\protect\citeauthoryear{Porredon, Crocce, Fosalba  et~al.}{Porredon et~al.}{2022}]{DES:2021bpo}
Porredon A.,  Crocce M.,  Fosalba P.,   et~al., 2022, \mn@doi [PRD] {10.1103/PhysRevD.106.103530}, 106, 103530

\bibitem[\protect\citeauthoryear{{Prince}, {Rogozenski}, {{\v{S}}ar{\v{c}}evi{\'c}}  \& {Gawiser}}{{Prince} et~al.}{2025}]{2025AAS...24525015P}
{Prince} H.,  {Rogozenski} P.,  {{\v{S}}ar{\v{c}}evi{\'c}} N.,   {Gawiser} E.,  2025, in American Astronomical Society Meeting Abstracts \#245. p. 250.15

\bibitem[\protect\citeauthoryear{Salcedo, Wibking, Weinberg  et~al.}{Salcedo et~al.}{2020}]{wl_degeneracies}
Salcedo A.~N.,  Wibking B.~D.,  Weinberg D.~H.,   et~al., 2020, \mn@doi [MNRAS] {10.1093/mnras/stz2963}, 491, 3061

\bibitem[\protect\citeauthoryear{Sipp, Sch{\"a}fer  \& Reischke}{Sipp et~al.}{2021}]{Sipp_2020}
Sipp M.,  Sch{\"a}fer B.~M.,   Reischke R.,  2021, \mn@doi [MNRAS] {10.1093/mnras/staa3710}, 501, 683

\bibitem[\protect\citeauthoryear{Takahashi, Sato, Nishimichi  et~al.}{Takahashi et~al.}{2012}]{halofit_takahashi}
Takahashi R.,  Sato M.,  Nishimichi T.,   et~al., 2012, \mn@doi [ApJ] {10.1088/0004-637X/761/2/152}, 761, 152

\bibitem[\protect\citeauthoryear{Tallada, Carretero, Casals  et~al.}{Tallada et~al.}{2020}]{TALLADA2020100391}
Tallada P.,  Carretero J.,  Casals J.,   et~al., 2020, \mn@doi [A\&C] {10.1016/j.ascom.2020.100391}, 32, 100391

\bibitem[\protect\citeauthoryear{Tanoglidis, Chang  \& Frieman}{Tanoglidis et~al.}{2020}]{photoz_galaxyclustering}
Tanoglidis D.,  Chang C.,   Frieman J.,  2020, \mn@doi [MNRAS] {10.1093/mnras/stz3281}, 491, 3535

\bibitem[\protect\citeauthoryear{Taylor, Kitching  \& McEwen}{Taylor et~al.}{2018}]{Taylor:2018nrc}
Taylor P.~L.,  Kitching T.~D.,   McEwen J.~D.,  2018, \mn@doi [PRD] {10.1103/PhysRevD.98.043532}, 98, 043532

\bibitem[\protect\citeauthoryear{Tutusaus, Martinelli, Cardone  et~al.}{Tutusaus et~al.}{2020}]{Tutusaus2020}
Tutusaus I.,  Martinelli M.,  Cardone V.~F.,   et~al., 2020, \mn@doi [A\&A] {10.1051/0004-6361/202038313}, 643, A70

\bibitem[\protect\citeauthoryear{Wang}{Wang}{2008}]{Wang:2008zh}
Wang Y.,  2008, \mn@doi [PRD] {10.1103/PhysRevD.77.123525}, 77, 123525

\bibitem[\protect\citeauthoryear{Wong, Brown, Duncan  et~al.}{Wong et~al.}{2025}]{wong_3x2pt_optimization}
Wong J. H.~W.,  Brown M.~L.,  Duncan C. A.~J.,   et~al., 2025, A\&A, submitted, arxiv:2501.07559

\bibitem[\protect\citeauthoryear{Wright, Stölzner, Asgari  et~al.}{Wright et~al.}{2025}]{Wright:2025xka}
Wright A.~H.,  Stölzner B.,  Asgari M.,   et~al., 2025, arxiv:2503.19441

\bibitem[\protect\citeauthoryear{Zuntz, Paterno, Jennings  et~al.}{Zuntz et~al.}{2015}]{Zuntz_2015}
Zuntz J.,  Paterno M.,  Jennings E.,   et~al., 2015, \mn@doi [A\&C] {10.1016/j.ascom.2015.05.005}, 12, 45

\bibitem[\protect\citeauthoryear{Zuntz, Lanusse, Malz  et~al.}{Zuntz et~al.}{2021}]{LSST_3x2pt_optimization}
Zuntz J.,  Lanusse F.,  Malz A.~I.,   et~al., 2021, \mn@doi [OJAp] {10.21105/astro.2108.13418}, 4, 13418

\makeatother
\end{thebibliography}

\appendix

\section{SOM implementation}\label{appendix:som}

In order to utilise the SOM to estimate photo-$z$'s, it must first be trained. The training starts with the initialisation of the SOM, where we define the weights of each voxel. The weights are set to random colour values using a Gaussian distribution centred at the average colour of the training sample galaxies. The training sample would consist of spectroscopic galaxies in a survey. Since we used a simulated catalogue our training sample contains the true redshift of the galaxies. The training of the SOM is carried out in a sequential manner by finding the voxel's weights that more closely match the training galaxy colours. For each training galaxy, the weights of the whole SOM are modified to more closely reflect the galaxy colour vector, $\vec{x}$. To find the best matching voxel to each galaxy we compute the distance between a given voxel's weights and a given galaxy as: 

\begin{equation}
    d_{jk} = \sum^{m}_{i} \frac{\left(x_i - w_{jki}\right)^2}{\sigma_i^2}\,,
\end{equation}
where $\sigma_i$ is the error in each colour measurement for a given galaxy. The voxel $jk$ whose distance $d_{jk}$ corresponds to the smallest value becomes the best-matching unit (BMU) for that particular galaxy. The weights of all the voxels in the map are updated. The amount of change is related to the distance from each voxel to the BMU for a given iteration, with closer voxels changing more. For each iteration in the training, the weights of each voxel in the map change in the following way:

\begin{equation}
    \vec{w}_{jk}(t+1) = \vec{w}_{jk}(t) + a(t)\, H_{j'k',jk}(t) \left(\vec{x}-\vec{w}_{jk}(t)\right)\,,
\end{equation}
where $j'k'$ are the indices of the BMU voxel, $t$ is the number of iterations of learning that have been carried out. Our choice was to have four iterations per galaxy in the training sample, as there was little to no improvement after training with a single iteration per galaxy already.\\

The weights of the map are all changed in each iteration with the goal of forming clusters based on the colour phenotype of galaxies. With $a(t)$ being a learning function that decreases the impact of each learning iteration as the number of iterations, $N_{iter}$, used in learning increases and a proximity parameter, $H_{j'k',jk}(t)$, that modulates the degree of change in the voxels surrounding the BMU for each iteration:

\begin{equation}
    a(t) =0.5^{2t/N_{iter}}\,,
\end{equation}

\begin{equation}
    H_{j'k',jk}(t) = e^{-D_{j'k',jk}^2/\sigma^2(t)}\,,
\end{equation}
with $D_{j'k',jk}$ being the distance between the position of the BMU at $j'k'$ and the voxel $jk$ on the map and $\sigma(t)$ modulating the diminishing effects of each iteration as a function of the distance between voxels on the map as:

\begin{equation}
    \sigma(t) = n_1 \left(\frac{1}{n_1}\right)^{2t/N_{iter}}\,,
\end{equation}
with $n_1$ being the shortest side of the map. The learning process of the SOM is such that the initial learning iterations will have a greater impact on the weights and also have such impact over a larger area of the map around the BMUs. As the training progresses and the iteration parameter $t$ increases, each new iteration will have a smaller impact on the weights and the impact will be limited to the BMU and a few voxels around it. In this way, colour phenotype clusters are first defined in the initial stages of the training and, as the iterations continue, the clusters become more differentiated with specific colour phenotypes.\\

The degree of completeness of the training sample will determine the regions in colour phenotype that will be covered by the SOM. This means that galaxies at high redshift, being less common, will be poorly characterised compared to those at lower redshift. Each voxel will have a redshift distribution composed of the true redshift of the galaxies assigned to that voxel during the training of the SOM. In order to reduce the number of catastrophic outliers, voxels with less than three redshifts in their $n(z)$'s are not used for the photo-$z$ estimation. We perform our learning process with $N_{iter} = 4\times10^5$ iterations, which is 4 times the training sample size, although using as many iterations as galaxies in the sample is enough according to \cite{rainbow}. Galaxies in the training sample will be cycled through in random order.\\

\begin{figure}
  \centering

  \begin{subfigure}{0.45\textwidth}
    \centering
    \includegraphics[width=\linewidth]{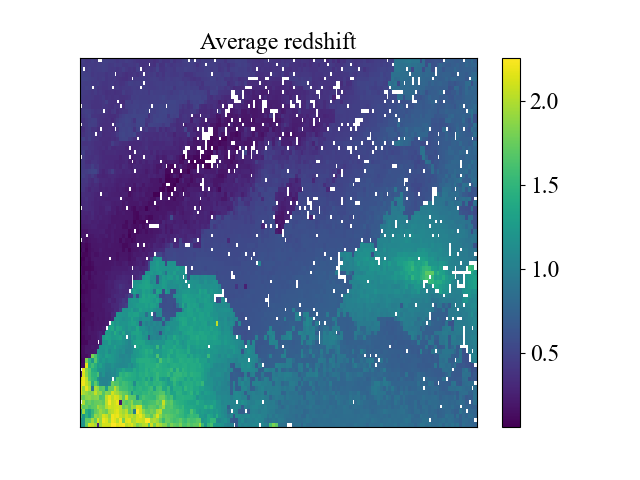}
    \caption{Redshift map of the SOM. The average redshift of the galaxies from the training sample is assigned to each voxel in the map. White voxels had no galaxies assigned during training.}
    \label{fig:redshiftmap}
  \end{subfigure}
  \hfill
  \begin{subfigure}{0.45\textwidth}
    \centering
    \includegraphics[width=\linewidth]{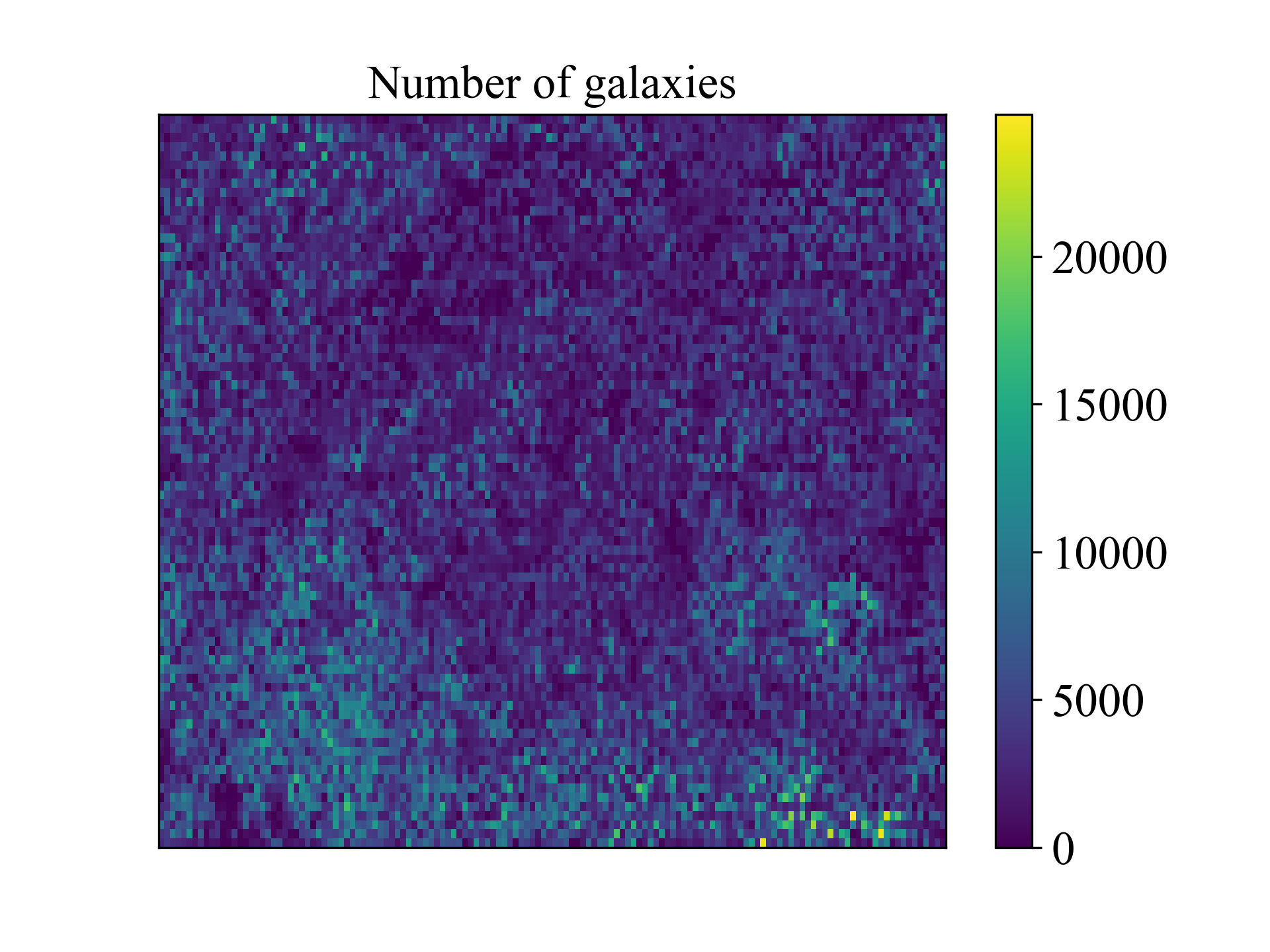}
    \caption{Galaxy density map of the SOM. The number of photo-z galaxies that are assigned to each voxel.}
    \label{fig:voxelmap}
  \end{subfigure}
    \caption{SOM visualisations: redshift map of the SOM after training and galaxy counts from the photo-$z$ galaxy catalogue.}
 \label{fig:SOM_maps}
\end{figure}

The size of the SOM in terms of its number of voxels is also relevant to the performance of the SOM. It must be large enough so that phenotype clusters can be formed but small enough to not be sparsely populated with many clusters. The former problem would lead to a poor characterisation of galaxies and the latter would be analogous to over-fitting. In addition, the shape of the SOM will also affect the formation of phenotype clusters. Rectangular shapes allow for more differentiated phenotype clusters to be formed during the training process. Our choice was to use a rectangular SOM of size 100 by 140 voxels, similar to the one in \cite{rainbow}, which is 75 by 150.\\

In Fig.\,\ref{fig:redshiftmap} we can see the resulting SOM. The redshift value of each voxel is determined during training by the average redshift of galaxies from the training sample that were assigned to that voxel. We show the resulting redshift map of the SOM, where redshift clusters appear. While some areas have smooth gradients in redshift, there are abrupt differences that account for similar colour phenotypes with substantially different redshift. This showcases the capacity of the SOM to distinguish colour phenotypes in order to avoid redshift degeneracies.\\

To generate the photo-$z$ assigned to each galaxy, a sample is generated from a normal distribution defined with the mean and standard deviation of the galaxies' redshift in each voxel. We have found this method preferable to directly using the voxel's $n(z)$ distribution as a probability distribution when estimating photo-$z$'s to reduce over-fitting. The resulting distribution of photo-$z$ galaxies on the SOM can be seen in Fig.\,\ref{fig:voxelmap}. The most populated areas of the SOM coincide with areas of medium redshift while high-redshift areas are less populated. Galaxies with similar colour phenotypes will be close on the map; having clusters of high and low redshift close on the map indicates the capacity of the SOM to differentiate galaxies at low and high redshift despite similar colour profiles. 

\section{Optimisation with equally-populated bins as initial point}\label{sec:equipop}

\begin{figure}
  \centering
    \includegraphics[width=\linewidth]{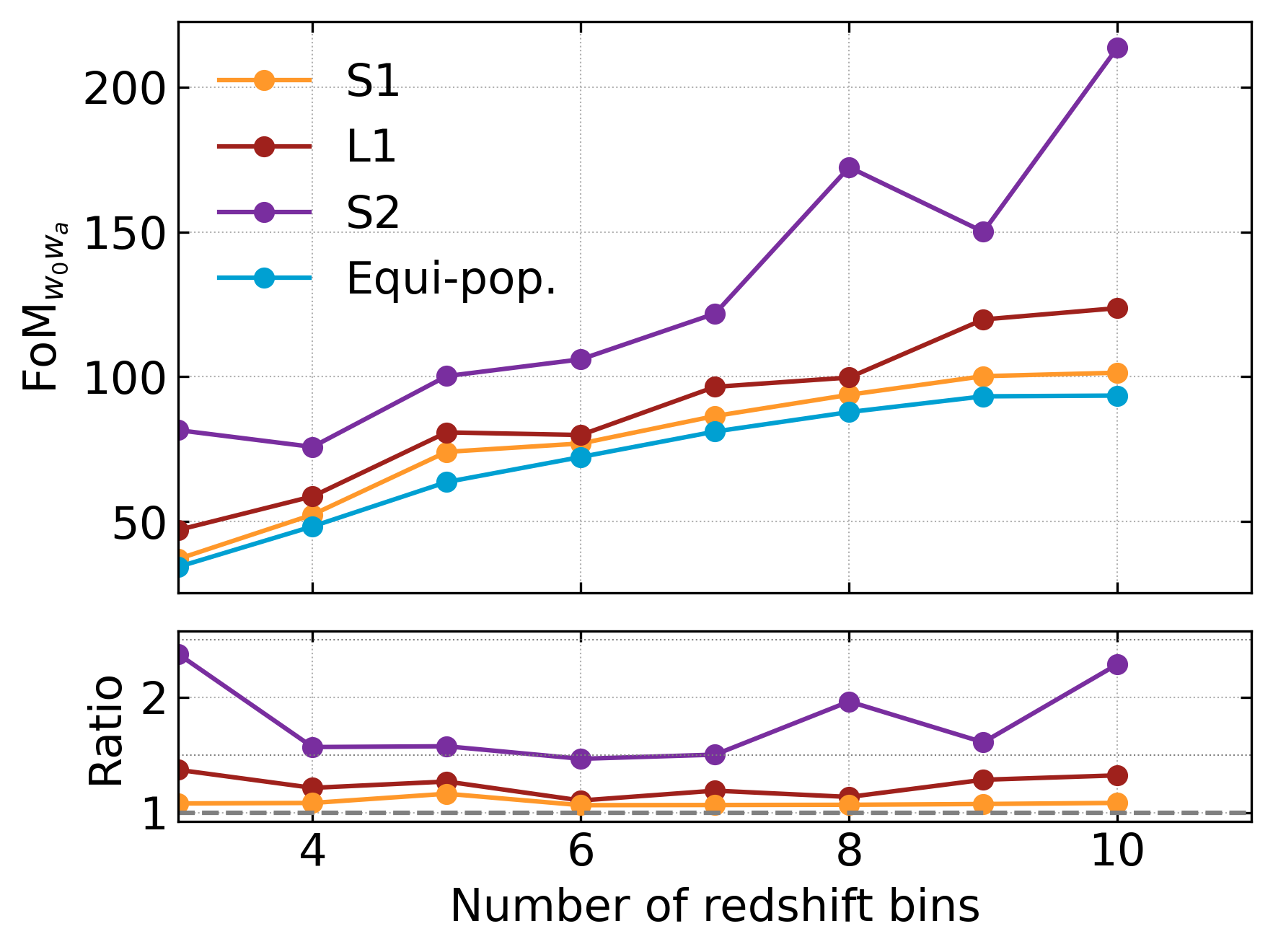}
    \caption{Optimisation starting from equally-populated redshift bins.}
 \label{fig:equi-populated optimization}
\end{figure}

For the sake of completeness, we have also run the optimisation algorithm with another set of initial conditions. Choosing equally-populated redshift bins instead of equal-width bins does not change the outcome of the optimisation. Looking at Fig.\,\ref{fig:equi-populated optimization} we can see how after applying the iterative optimisation process in three instances we already recover an improvement in the FoM for $w_{\rm 0} w_a$ of a factor above two. This is expected because the method has the freedom to explore tomographic configurations well beyond the initial conditions. 

\section{Simultaneous optimisation of source and lens samples}\label{appendix: simultaneous optimization}

\begin{figure}
  \centering
    \includegraphics[width=\linewidth]{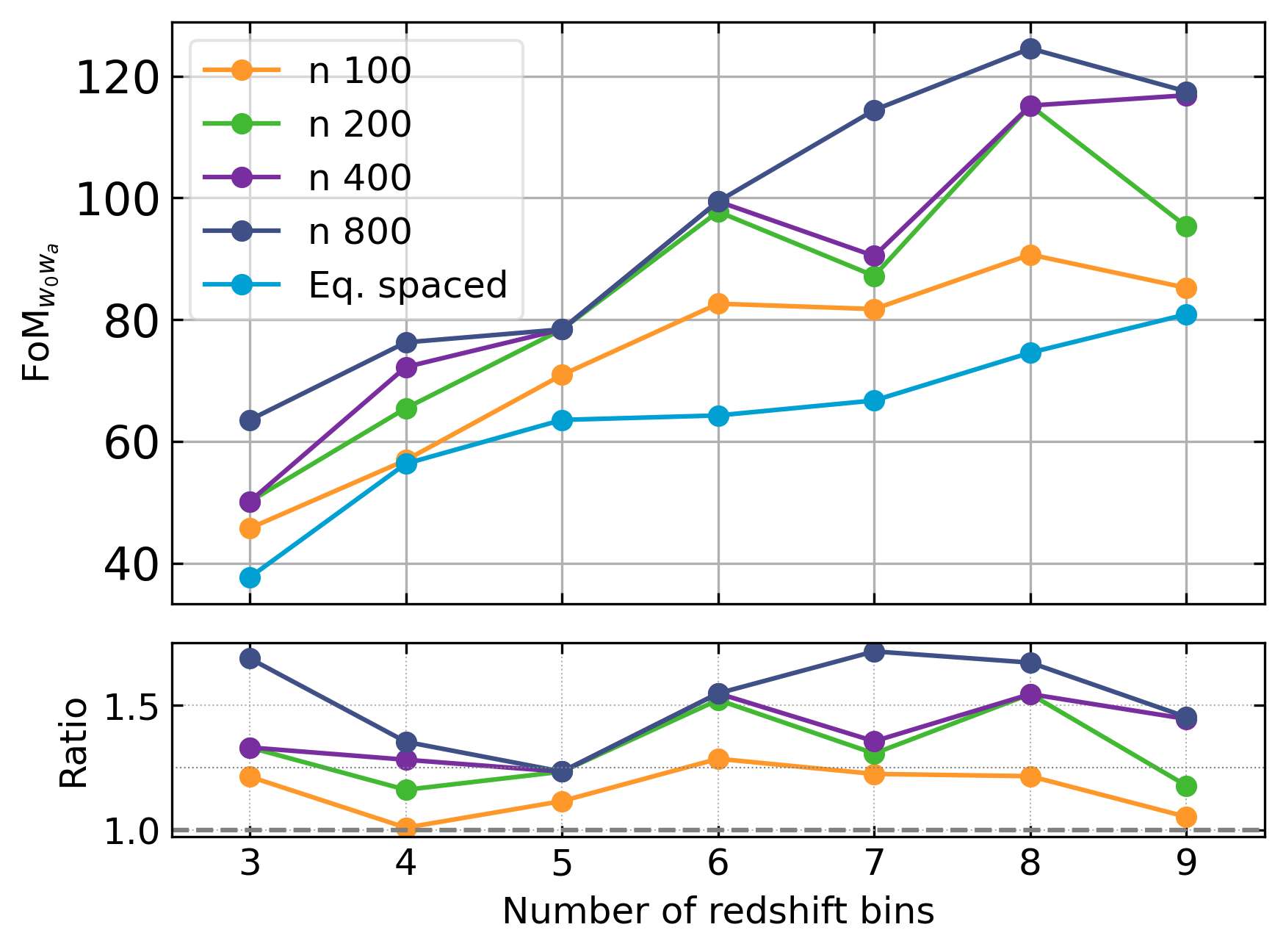}
    \caption{Simultaneous optimisation of the source and lens samples. Each line corresponds to the FoM of the best-performing tomography when randomly sampling both the source and lens tomographic configurations. The parameter 'n' corresponds to the total number of runs, with n=800 being the same number of runs needed for the 4 iterations of the iterative optimisation.}
 \label{fig:simultaneous optimization}
\end{figure}

We see in Fig.\,\ref{fig:simultaneous optimization} the results of using a simpler approach where we compute the FoM varying the $n(z)$ of both the source and lens tomography simultaneously. Using this method instead of varying the $n(z)$ of sources for a fixed lens configuration and vice-versa is much less efficient in reaching an optimised tomography. This is because the optimal $n(z)$ of the source galaxies depends on the lens tomography, hence our iterative method can better optimise the samples taking this interplay between lens and source galaxies into account. Thus, the sampling needs to be much larger to converge into an optimal tomography for the source and lens galaxies. The result is that the improvement in the FoM is not as smooth or predictable when we increase the sampled configurations by the same amount as what would correspond to S1, L1, and L2 in the iterative optimisation method. This showcases the advantage of using the iterative method of alternating between source and lens optimisation with respect to sampling both the source and lens tomographic configurations simultaneously.

\section{Optimised likelihood contours}\label{appendix: likelihoods}

\begin{figure*}
    \centering
     \includegraphics[width=1\textwidth]{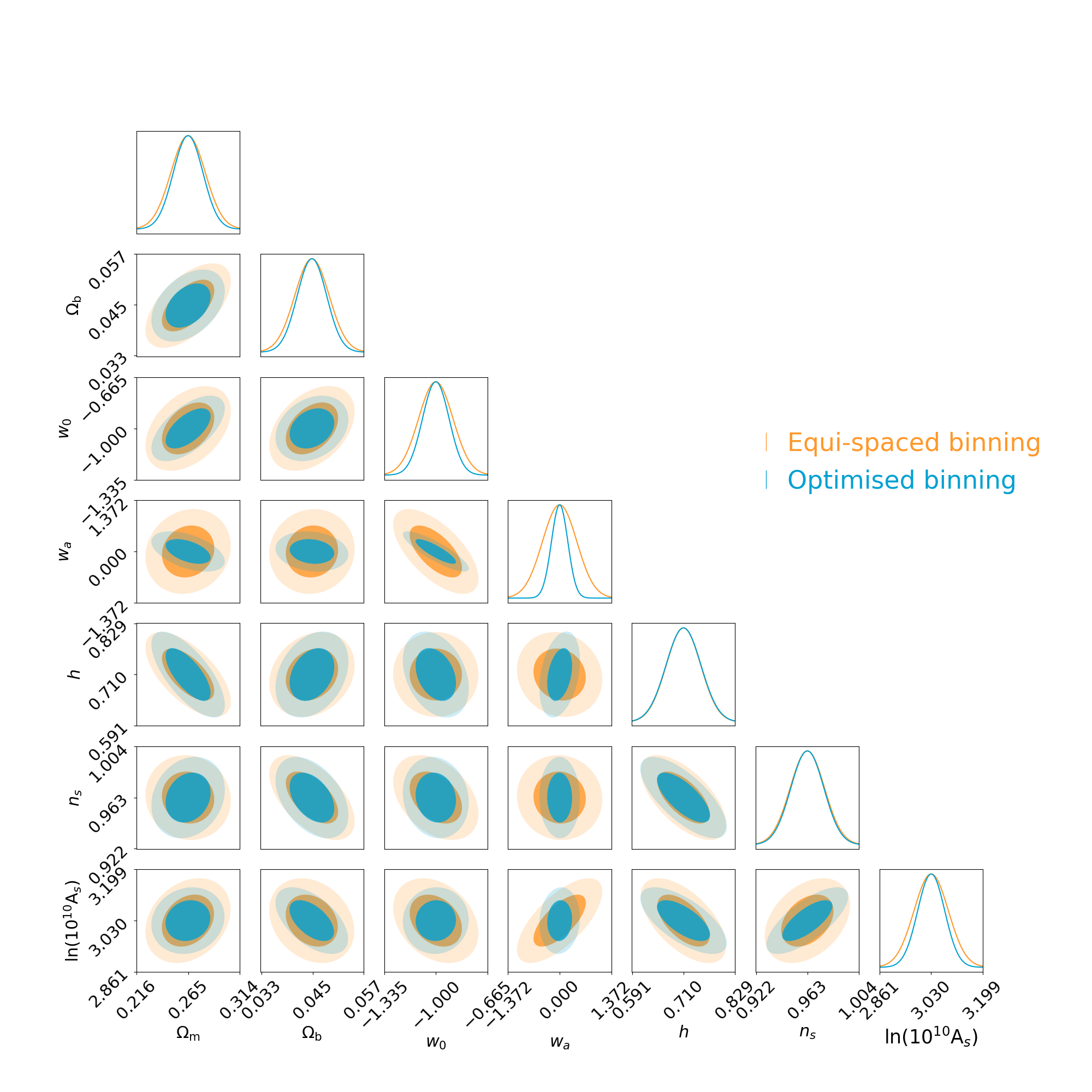}
 \caption{Likelihood contours at 1-$\sigma$ and 2-$\sigma$ estimated through the Fisher matrix for the 7 cosmological parameters of the $w_{\rm 0} w_a\text{CDM}$ model. Plotted are the contours obtained with the optimised tomography for $w_{\rm 0}$ $w_a$ (blue) and with the equally-spaced tomography (orange) for 8 redshift bins.}
 \label{fig:fisher_3x2pt_w0wa}
  \end{figure*}

In Fig.\,\ref{fig:fisher_3x2pt_w0wa} we show the constraints for the $w_{\rm 0} w_a\text{CDM}$ model parameters in the base case (equally-spaced tomography) and the optimised sample selection case for 8 tomographic redshift bins. Our optimisation method has affected the constraints for $w_a$ most dramatically while providing more modest improvements for other cosmological parameters like $w_{{\rm 0}}$ or $\text{A}_{\rm s}$, while other parameters are mostly unaffected. Crucially, our tomographic optimisation does not degrade the constraints for other parameters of the model even when the target was $\text{FoM}_{w_{\rm 0} w_a}$.

\end{document}